\documentclass{iopart}
\usepackage{euscript,iopams,amssymb,amsfonts,graphicx,bm}
\usepackage{pgfplots,setstack}

\usepackage{float}
\usepackage{epsfig}
\usepackage{esint}
 \usepackage{tikz-cd}

\usepackage{bm,braket}

\eqnobysec
\newcommand{\calU}{{\mathcal U}}

\newcommand{\calQ}{{\mathfrak Q}}

\newcommand{\calF}{{\mathcal F}}

\newcommand{\calM}{{\mathcal M}}
\newcommand{\calC}{{\mathcal C}}

\newcommand{\R}{{\mathbb R}}
\renewcommand{\L}{{\mathbb L}}

\renewcommand{\P}{\mathbb{P}}

\newcommand{\p}{\widetilde{p}}
\newcommand{\J}{\widetilde{J}}
\newcommand{\Q}{\widetilde{Q}}
\newcommand{\G}{\widetilde{G}}
\newcommand{\x}{\mathbf{x}}
\newcommand{\y}{\mathbf{y}}
\renewcommand{\e}{{\mathrm e}}
\newcommand{\E}{{\mathbb E}}
\newcommand{\pcb}[1]{\textcolor{black}{#1}}
\newcommand{\n}{\mathbf n}
\renewcommand{\a}{\mathbf a}
\newcommand{\z}{\mathbf z}
\newcommand{\calT}{{\mathcal T}}
\renewcommand{\P}{\mathbb P}

\newcommand{\q}{\widetilde{q}}
\newcommand{\f}{\widetilde{f}}
\newcommand{\F}{\widetilde{F}}

\renewcommand{\S}{\widetilde{S}}

\newcommand{\wrho}{\widetilde{\rho}}

\begin{document}

 \title[Renewal theory for Brownian motion with stochastically gated targets]{Renewal theory for Brownian motion with stochastically gated targets}

\author{Paul C. Bressloff}
\address{Department of Mathematics, Imperial College London, 
London SW7 2AZ, UK}
\ead{p.bressloff@imperial.ac.uk}

\begin{abstract} 
There are a wide range of first passage time \pcb{(FPT)} problems in the physical and life sciences that can be modelled in terms of a Brownian particle binding to a reactive surface (target) and initiating a downstream event (absorption). However, prior to absorption, the particle may undergo several rounds of surface attachment (adsorption), detachment (desorption) and diffusion. Alternatively, the surface may be stochastically gated so that absorption can only occur when the gate is open. In both cases one can view each return to the surface as a renewal event. In this paper we develop a renewal theory for stochastically gated FPT problems along analogous lines to previous work on adsorption/desorption processes. We proceed by constructing a renewal equation that relates the joint probability density for particle position and the state of a gate (or multiple gates) to the probability density and FPT density for a totally absorbing (non-gated) boundary. This essentially decomposes sample paths into an alternating sequence of bulk diffusion and instantaneous adsorption/desorption events, which is terminated when adsorption coincides with an open gate. In order to ensure that diffusion restarts in a state that avoids immediate re-adsorption. we assume that whenever the particle reaches a closed boundary it is instantaneously shifted a distance $\epsilon$ from the boundary (boundary-induced resetting). We explicitly solve the renewal equation for $\epsilon>0$ and show how the solution to the original gated FPT problem is recovered in the limit $\epsilon\rightarrow 0$. Through a series of examples, we who how renewal theory proves a general mathematical framework for modelling stochastically gated targets. \end{abstract}

\maketitle
\section{Introduction} 

\begin{figure}[b!]
\centering
\includegraphics[width=12cm]{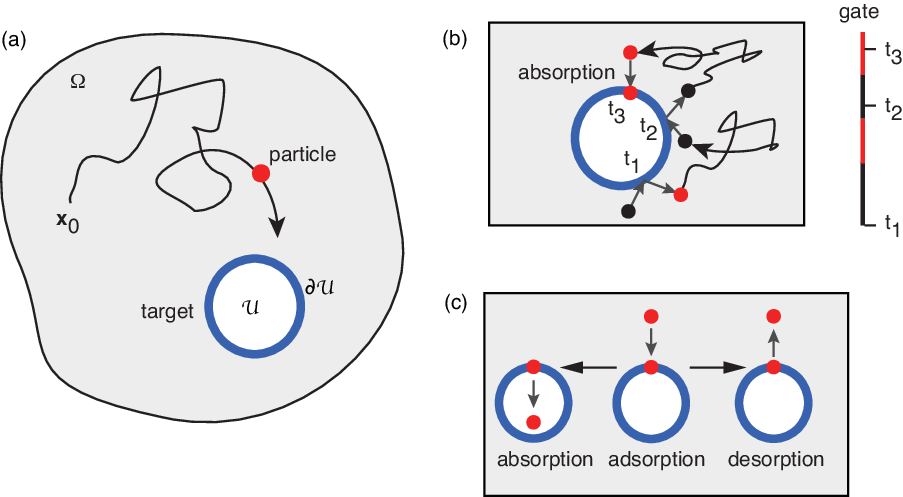} 
\caption{a) Single-particle diffusion in a bounded domain $\Omega$ containing a target $\calU\subset \Omega$. The exterior boundary $\partial \Omega$ is totally reflecting whereas the target boundary $\partial \calU$ is absorbing. (b) Stochastically gated adsorption. The particle randomly switches between a reactive and non-reactive state, and can only be absorbed at $\partial \calU$ when in the reactive state otherwise it is reflected. Equivalently, the surface $\partial \calU$ switches between an open (absorbing) state and a closed (reflecting) state. (c) When a particle reaches the surface it may temporarily bind to $\partial \calU$ (adsorb) and then either unbind and restart bulk diffusion (desorb) or be permanently removed (absorb). }
\label{fig1}
\end{figure}

A major topic in single-particle diffusion is the first passage
time (FPT) problem for a sample path to reach a target $\calU\subset \Omega$ (or a set of targets) within a bounded domain $ \Omega \subset \R^d$ \cite{Grebenkov24B}, see Fig. \ref{fig1}(a). This is often modelled in terms of Brownian motion (BM) in $\Omega\backslash \calU$ supplemented by the stopping condition that the process is terminated as soon as the particle reaches the target boundary $\partial \calU$. In other words, $\partial \calU$ is treated as a totally absorbing surface. (Note that the exterior boundary $\partial \Omega$ could also be an absorbing target, and there could be multiple interior targets. Here $\partial \Omega$ is assumed to be totally
reflecting.) Absorption may represent a particle binding to the surface and triggering a downstream event such as a chemical reaction or the transfer of resources between the particle and target. Examples include an extracellular signalling molecule (ligand) binding to a receptor in the surface membrane of a cell \cite{Bressloff22B}, and a foraging animal finding a target containing food or shelter \cite{Bell91,Bartumeus09,Viswanathan11}. 
However, the assumption of instantaneous (total) absorption neglects the possibility that when the particle reaches the target surface, additional constraints must be satisfied before the stochastic process is terminated. 

First, physical properties of a surface may make it difficult for a particle to bind to the surface on its first attempt, thus
requiring an alternating sequence of bulk diffusion interspersed with local surface interactions prior to binding. In other words, the reactive surface $\partial \calU$ is partially absorbing \cite{Szabo84,Schumm21}. If the rate of absorption $\kappa_0$ is constant, then the corresponding probability density $p(\x,t)$ for particle position satisfies a Robin boundary condition on $\partial \calU$. The totally absorbing case is recovered in the limit $\kappa_0\rightarrow \infty$. More generally, the probability of absorption may depend non-trivially on the amount of surface-particle contact time, as considered in encounter-based models of diffusion-mediated absorption  \cite{Grebenkov20,Grebenkov22,Bressloff22,Bressloff22a,Grebenkov24}. A second mechanism for non-instantaneous binding is stochastic gating, see Fig. \ref{fig1}(b). A classical example in chemical reaction kinetics is a diffusing enzyme that stochastically switches between a reactive state that allows it to bind to the substrate and a non-reactive state that prevents binding. Alternatively, the substrate itself could be gated as in the case of certain protein receptors in biological cells. These examples underly the more general problem of stochastically gated chemical reactions \cite{Budde95,Spouge96,Sheu97,Sheu99,Shin18,Boyer19,Boyer21,Boyer21a,Scher21,Scher21a,Kumar23,Scher24}. Stochastically gated diffusion also plays a role in cellular transport processes \cite{Bressloff22B}. For example, 
the random opening and closing of channels and pores in the plasma membrane of a cell or subcellular compartment affects the diffusive exchange of molecules between the interior and exterior of the cell or compartment \cite{Reingruber10,PCB15a,PCB15b,Godec17}. A related example concerns the intercellular transport of molecules through cells coupled by stochastically gated gap junctions \cite{Bressloff16a,Bressloff16b}, whereas an example at the organismal level is oxygen transport in insect respiration \cite{Lawley15,Berez16}. In all of these cases, termination of the stochastic process corresponds to exit of a particle through an open gate rather than binding to a reactive surface.

A third constraint is that even if a particle immediately binds upon first contact to a surface, it may subsequently unbind and return to diffusion in the bulk (desorption) before initiating a downstream reaction or being permanently removed from the system, see Fig. \ref{fig1}(c). The target surface is effectively ``sticky.'' In such cases, the binding event is distinct from the termination event so we refer to the former as adsorption and the latter as absorption. Reversible or partially reversible surface adsorption/desorption processes have been studied for many years in physical chemistry, see for example Refs. \cite{Baret68,Agmon84,Adam87,Adam87a,Agmon89,Agmon90,Agmon93,Franses95,Passerone96,Reuveni23,Scher23,Grebenkov23,Bressloff25a}. A common example in cell biology is
reversible ligand-receptor binding in cell signalling, with the absorption of the ligand-receptor complex into the cell interior requiring an active process known as endocytosis. Analogously, a foraging animal that finds a target may fail to access the resources within the interior of the target (absorption), and thus leave the target (desorption) to continue its search \cite{Bressloff25b}.

One consequence of a sticky boundary $\partial \calU$ is that a Brownian particle typically undergoes several rounds of bulk diffusion followed by first-return to $\partial \calU$ prior to a final absorption event. Each first-return to $\partial \calU$ can be viewed as a renewal of the stochastic process. This has motivated the construction of renewal equations that relate the probability density and FPT density for absorption to the corresponding quantities for adsorption, which are usually simpler to compute \cite{Grebenkov23,Scher23,Bressloff25a,Bressloff25b}. The renewal equations take the form of implicit integral equations that can be solved explicitly using Laplace transforms and the convolution theorem. However, a crucial step in the analysis is ensuring that BM restarts in a state that avoids immediate re-adsorption. This can be achieved by taking the sticky surface to be partially adsorbing \cite{Grebenkov23,Scher23,Bressloff25a} or by resetting the particle to its initial position away from the wall following desorption \cite{Bressloff25b}. This form of desorption-induced stochastic resetting is distinct from spontaneous resetting in the bulk domain \cite{Evans20}. The renewal approach provides a general framework for incorporating more complicated models of adsorption and desorption. First, one can consider a non-Markovian model of desorption by taking the waiting time density for the duration of an adsorbed state to be non-exponential. Second, one can take the target surface to be partially rather than totally adsorbing and apply encounter-based methods \cite{Grebenkov23,Bressloff25a}. Third, one can include both spontaneous stochastic resetting in the bulk and desorption-induced resetting \cite{Bressloff25b}.

In this paper we develop an analogous renewal theory for stochastically gated diffusion processes. We begin by considering BM on the half-line $[0,\infty)$ with a stochastically gated boundary at $x=0$ (section 2). We first solve the forward Kolmogorov equation for the probability densities $\rho_k(x,t)=\E[\rho(x,t)\delta_{k,\sigma(t)}]$, where $\rho(x,t)$ is the marginal probability density for particle position anmd expectations are taken with respect to realisations of the gating process. We then construct a first renewal equation that relates $\rho_k(x,t)$ to the corresponding probability density $p(x,t)$ for BM on the half-line with a totally absorbing boundary condition at $x=0$. In the renewal formulation, we treat each particle encounter with the boundary $x=0$ when the gate is closed as an instantaneous adsorption/desorption event. That is,  sample paths are effectively decomposed into an alternating sequence of bulk diffusion and instantaneous adsorption/desorption events, which is terminated when adsorption coincides with an open gate. As in the case of sticky targets, we need to restart the diffusion process in a state that avoids immediate re-adsorption. Therefore, we assume that following adsorption at $x=0$ when the gate is closed, the particle immediately restarts at a distance $\epsilon$ from the origin. The latter can be interpreted as a form of so-called boundary-induced resetting \cite{Bressloff25b}. We then obtain an explicit solution of the  renewal equation for $\epsilon>0$ using Laplace transforms and the convolution theorem.  We then show how the resulting solution in the limit $\epsilon\rightarrow 0$ is equivalent to the solution of the Kolmogorov equation. However, the renewal equation provides a more general mathematical framework for modelling a stochastically gated boundary by explicitly separating the FPT problem for reaching the detecting the boundary (absorption)  and the subsequent rule for restarting diffusion. In section 3, we apply the renewal method to a pair of 1D FPT problems. First, we calculate the MFPT for absorption of a Brownian particle with stochastic resetting on the half-line and a stochastically gated boundary.  Second, we consider the FPT problem for single-particle diffusion in the finite interval with a stochastically gated boundary at each end. If the particle switches between a reactive and non-reactive state then both ends are either totally absorbing or totally reflecting (totally correlated boundary conditions). On the other hand, if the boundaries themselves are stochastically gated, then they independently switch between open and closed states (uncorrelated boundary conditions). In section 4, we consider a natural extension of the 1D problems analysed in section 3, namely, diffusion in a domain $\Omega \subset \R^d $, $d>1$, with a stochastically gated target surface $\partial \Omega \subset \R^{d-1}$. The corresponding renewal equation now involves a spatial integral with respect to $\y \in \partial \Omega$, which complicates the analysis. Nevertheless, we show how the Laplace transformed renewal equation can be solved by extending spectral methods previously developed in Refs. \cite{Grebenkov20,Bressloff22a,Grebenkov23,Bressloff25a}. In particular, taking the limit $\epsilon \rightarrow 0$ naturally leads to the appearance of a Dirichlet-to-Neumann (D-to-N) operator on the target surface $\partial \Omega$. We illustrate the theory using the example of diffusion in a sphere.

We end this introduction by noting that a complementary renewal formulation of stochastically gated processes has recently been developed by Scher et al. \cite{Scher21,Scher21a,Kumar23,Scher24}. In particular, in Ref. \cite{Scher24} these authors focus on a stochastically gated trap $\calU$ in 1D rather than a point-like boundary $\partial \calU$ -- a diffusing particle can freely enter and exit the trap and is absorbed within $\calU$ as soon as the gate opens. Instead of constructing integral renewal equations for the full propagator, they derive recursive equations relating the FPT for absorption to the corresponding FPT density for adsorption in the absence of gating  by conditioning on whether the particle is located within or outside the trap when the gate reopens. This allows the authors to obtain exact expressions for the Laplace transform of the FPT density for a 1D trap. In the discussion, we briefly indicate some of the difficulties in extending our renewal formulation to the trapping problem. 

\section{Brownian particle on the half-line with a stochastically gated boundary}

In this section we introduce the renewal formulation by considering Brownian motion on the half-line $[0,\infty)$ with a stochastically gated boundary at $x=0$. We first use Laplace transforms to solve the forward Kolmogorov equation for the joint propagator $\rho$ that tracks both particle position and the state of the gate, see also Refs. \cite{Boyer19,Boyer21,Boyer21a}. We then construct the corresponding renewal equation that relates the joint propagator $\rho$ to the marginal propagator for particle position in the case of a totally absorbing (non-gated) boundary. We also assume that the particle instantaneously resets to a position $\epsilon>0$ whenever it is reflected off a closed gate. This allows us to solve the renewal equation using Laplace transforms and the convolution theorem. We then show how the solution of the forward Kolmogorov equation is recovered in the limit $\epsilon \rightarrow 0$.

  \subsection{Forward Kolmogorov equation} 
  
  \begin{figure}[b!]
  \centering
  \includegraphics[width=12cm]{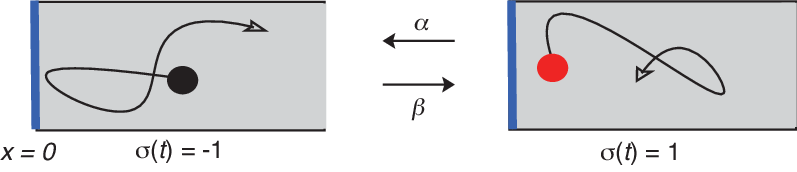}
  \caption{Brownian particle in the half-line $[0,\infty)$ with a stochastically gated boundary at $x=0$. The gate is open if $\sigma(t)=1$ and closed if $\sigma(t)=-1$. Transitions $\sigma(t)\rightarrow -\sigma(t)$ are given by a two-state Markov chain with transition rates $\alpha,\beta$. The gate could represent a conformational state of the particle (shown) or the boundary. (We consider a 2D domain for illustrative purposes.)}
  \label{fig2}
\end{figure}

  Consider a Brownian particle diffusing in the interval $x\in [0,\infty)$ with a stochastically opening and closing gate at $x=0$, see Fig. \ref{fig2}. Gating could be determined either by the physical state of the particle or the boundary. Let $X(t)\in [0,\infty)$ denote the position of the particle at time $t$ and $\sigma(t)\in \{1,-1\}$ the corresponding state of the gate. The gate is open if $\sigma(t)=1$ and closed otherwise. The FPT for absorption is thus defined as
\begin{equation}
\calT=\inf\{t>0,\ X(t)=0,\sigma(t)=1\}.
\end{equation}
The discrete process $\sigma(t) $ evolves according to a two-state Markov chain with transition matrix\begin{equation}
{\bf K}=\left (\begin{array}{cc} 0& \beta \\ \alpha & 0 \end{array} \right ).
\label{axe2}
\end{equation}
 If $\Pi_{kk_0}(t)=\P[\sigma(t)=k|\sigma(0)=k_0]$, then the master equation for $\sigma(t)$ takes the form
\[\frac{d\Pi_{k k_0}}{dt}=\sum_{m=\pm 1}\bigg [K_{km}-\delta_{k,m}\sum_{l=\pm 1} K_{lk} \bigg ]\Pi_{mk_0}.\]
Using the fact that $\Pi_{-1k_0}(t)+\Pi_{1k_0}(t)=1$ we can solve this pair of equations to give
\begin{equation}
\label{Pi}
\Pi_{kk_0}(t)=\delta_{k,k_0}\e^{-t/\tau_c}+\sigma_k (1-\e^{-t/\tau_c}),\quad \tau_c=\frac{1}{\alpha+\beta},
\end{equation}
with $\sigma_1=\beta/(\alpha+\beta) $ and $\sigma_{-1}=\alpha/(\alpha+\beta) $.
Here $\tau_c$ is the relaxation time such that $\Pi_{kk_0}(t)\rightarrow \sigma_k$ in the limit $t\rightarrow \infty$.
Let $\rho_{k|k_0}(x,t|x_0)$ denote the full propagator for the pair $(X(t),\sigma(t))$ with
\[\fl \rho_{k|k_0}(x,t|x_0)dx=\P[x<X(t)<x+dx,\sigma(t)=k|X(0)=x_0,\sigma(0)=k_0]\]
and the initial condition $\rho_{k|k_0}(x,0|x_0)=\delta(x-x_0)\delta_{k.k_0}$.
The propagator evolves according to the forward Kolmogorov equation
\numparts
\begin{eqnarray}
\label{1Da}
\fl \frac{\partial \rho_{1|k_0}(x,t|x_0)}{\partial t}&=D\frac{\partial^2\rho_{1|k_0}(x,t|x_0)}{\partial x^2}-\alpha  \rho_{1|k_0}(x,t|x_0)+\beta \rho_{-1|k_0}(x,t|x_0), \\
\fl \frac{\partial \rho_{-1|k_0}(x,t|x_0)}{\partial t}&=D\frac{\partial^2\rho_{-1|k_0}(x,t|x_0)}{\partial x^2}+\alpha  \rho_{1|k_0}(x,t|x_0)-\beta \rho_{-1|k_0}(x,t|x_0)
\label{1Db}
\end{eqnarray}
for $x>0$ and
\begin{eqnarray}
\fl D\frac{\partial \rho_{-1|k_0}(0,t|x_0)}{\partial x}&=0,\quad \rho_{1|k_0}(0,t|x_0)=0.
\label{1Dc}
\end{eqnarray}
\endnumparts
We interpret the boundary condition at $x=0$ as follows. Whenever the particle hits the boundary either the gate is closed and the particle is reflected or the gate is open and the particle is absorbed. 
Introduce the survival probability
\begin{equation}
\label{S}
\calQ_{k_0}(x_0,t)=\int_0^{\infty}[\rho_{1|k_0}(x,t|x_0)+\rho_{-1|k_0}(x,t|x_0)]dx.
\end{equation}
Differentiating with respect to time $t$ and using equations (\ref{1Da})-(\ref{1Dc}) shows that the corresponding FPT density is
\begin{equation}
\label{dS}
\calF_{k_0}(x_0,t)\equiv -\frac{d\calQ_{k_0}(x_0,t)}{dt}=D\frac{\partial \rho_{1|k_0}(0,t|x_0)}{\partial x}.
\end{equation}
That is, $\calF_{k_0}(x_0,t)$ equals the net flux into the boundary from the bulk when the gate is open.

Laplace transforming equations (\ref{1Da})--(\ref{1Dc}) gives
\numparts
\begin{eqnarray}
\label{1DLTa}
\fl & D\frac{\partial^2\wrho_{1|k_0}(x,s|x_0)}{\partial x^2} -(\alpha +s)\wrho_{1|k_0}(x,s|x_0)+\beta  \wrho_{-1|k_0}(x,s|x_0)= -\delta(x-x_0)\delta_{1,k_0}, \\
 \label{1DLTb}
 \fl & D\frac{\partial^2\wrho_{-1|k_0}(x,s|x_0)}{\partial x^2} -(\beta +s)\wrho_{-1|k_0}(x,s|x_0)+\alpha  \wrho_{1|k_0}(x,s|x_0)\nonumber \\
 \fl & \hspace{5cm} = -\delta(x-x_0)\delta_{-1,k_0} \end{eqnarray}
\newpage  
\noindent for $x>0$ and
 \begin{eqnarray}
\label{1DLTc}
 &D\frac{\partial \wrho_{-1|k_0}(0,s|x_0)}{\partial x}=0,\quad \wrho_{1|k_0}(0,s|x_0)=0.
\end{eqnarray}
\endnumparts
Adding equations (\ref{1DLTa}) and (\ref{1DLTb}), and setting 
\begin{equation}
q(x,s)=\wrho_{1|k_0}(x,s|x_0)+\wrho_{-1|k_0}(x,s|x_0), 
\end{equation}
we have  
\begin{eqnarray}
 & D\frac{\partial^2q(x,s)}{\partial x^2} -sq(x,s)= -\delta(x-x_0), \quad x>0. \end{eqnarray}
Take the boundary condition for $q$ to be $ q(0,s)=A(s)$ with $A(s)$ to be determined self-consistently. It follows that
\begin{eqnarray}
\label{q}
q(x,s)&=\G(x,s|x_0)+\e^{-\sqrt{s/D}x} A(s),
\end{eqnarray}
where
\begin{equation}
\label{DG}
\G(x,s|x_0)=\frac{1}{2\sqrt{sD}}\bigg (\e^{-\sqrt{s/D}|x-x_0|}-\e^{-\sqrt{s/D}(x+x_0)}\bigg )
 \end{equation}
 is the Dirichlet Green's function of the modified Helmholtz equation on the half-line. Given the solution $q(x,s)$ we set 
   \begin{eqnarray}
 \fl  \wrho_{1|k_0}(x,s|x_0)=\sigma_{1}q(x,s)+j_{k_0}(x,s),\quad 
 \wrho_{-1|k_0}(x,s|x_0)=\sigma_{-1}q(x,s)-j_{k_0}(x,s),
 \end{eqnarray}
 with $j_{k_0}(0,s)=B(s)$.
 Substituting into equation (\ref{1DLTa}) implies that
 \begin{eqnarray}
 \fl & D\frac{\partial^2j_{k_0}(x,s)}{\partial x^2} -(\alpha+\beta +s)j_{k_0}(x,s)=-\delta(x-x_0)\bigg [\delta_{1,k_0}-\sigma_{1}\bigg ],\quad x>0.
 \end{eqnarray}
 The solution has the general form
 \begin{eqnarray}
 \label{j}
\fl  j_{k_0}(x,s)=  k_0\sigma_{-k_0}\G(x,s+\alpha+\beta|x_0)- \e^{-\sqrt{(s+\alpha+\beta)/D}x} B(s), \quad  x>0.
\end{eqnarray}

We have two unknown coefficients $A(s),B(s)$, which also depend on $(x_0,k_0)$, and two linearly independent boundary conditions given by (\ref{1DLTc}). Expressing the latter in terms of $q$ and $j_{k_0}$, we have
 \begin{eqnarray}
\left.  \sigma_{-1}\frac{\partial q(x,s)}{\partial x}\right |_{x=0}=\left .\frac{\partial j_{k_0}(x,s)}{\partial x}\right |_{x=0},\quad
   \sigma_{1}q(0,s)+j_{k_0}(0,s)= 0.
 \end{eqnarray}
 Combining with the general solutions (\ref{q}) and (\ref{j}) gives $B(s)=\sigma_1A(s)$ with
  \begin{eqnarray}
  \label{cond}
 A(s)=\frac{\sigma_{-1}\e^{-\sqrt{s/D}x_0}- k_0\sigma_{-k_0}\e^{-\sqrt{(s+\alpha+\beta)/D}x_0}}{\sigma_{-1}\sqrt{sD}+\sigma_1\sqrt{(s+\alpha+\beta)D}}.
  \end{eqnarray}
 In summary, given these coefficients, the solution of equations (\ref{1DLTa})-- (\ref{1DLTc}) is
\begin{eqnarray}
\fl  &\wrho_{k|k_0}(x,s|x_0) =\sigma_{k}\bigg [\G(x,s|x_0)+\e^{-\sqrt{s/D}x}A(s)\bigg ]\nonumber \\
\fl & \hspace{2cm}   +k\bigg [k_0\sigma_{-k_0}\G(x,s+\alpha+\beta|x_0)-\sigma_1\e^{-\sqrt{[s+\alpha+\beta]/D}x}A(s)\bigg ].
\label{sol}\end{eqnarray}

  \subsection{Renewal equation}

\begin{figure}[t!]
\centering
\includegraphics[width=10cm]{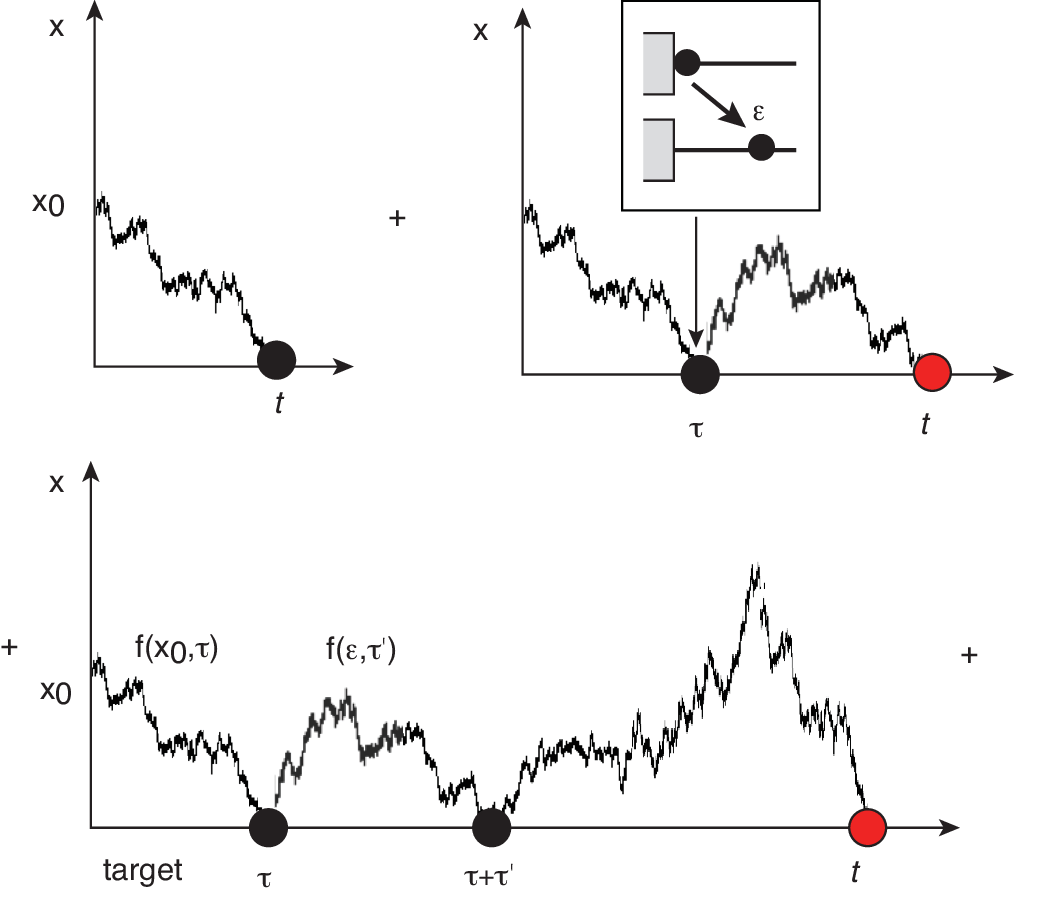} 
\caption{Example sample paths arising in the first three terms of the iterative expansion of the renewal equation (\ref{itrenewal}). (The time axis in successive diagrams is rescaled for illustrative purposes.)  Each black (red) circle represents hitting the boundary when the gate is closed (open). If the gate is closed then the reflected particle is infinitesimally displaced by an amount $\epsilon$ (see insert).}
\label{fig3}
\end{figure}

We now introduce a renewal formulation of the BVP for a stochastic gate at $x=0$ under the assumption that whenever the particle reflects off the closed gate it resets to a position $x=\epsilon$ (instantaneous adsorption/desorption protocol). The corresponding propagator $\rho^{(\epsilon)}_{kk_0}(x,t|x_0)$ satisfies a first renewal equation  of the form (for $x_0>0$)
 \begin{eqnarray}
  \label{renewal}
\fl  \rho^{(\epsilon)}_{k|k_0}(x,t|x_0)&=\Pi_{kk_0}(t)p(x,t|x_0)+\int_0^t  \rho^{(\epsilon)}_{k|-1}(x,t-\tau|\epsilon)\Pi_{-1,k_0}(\tau)  f(x_0,\tau)d\tau ,\end{eqnarray}
 with $p(x,t|x_0)$ the propagator for a totally adsorbing boundary:
\numparts
\begin{eqnarray}
\label{p1Da}
 \frac{\partial p(x,t|x_0)}{\partial t}&=D\frac{\partial^2p(x,t|x_0)}{\partial x^2},\quad 0<x<\infty,\\
p(0,t|x_0)&=0,\quad p(x,0|x_0)=\delta(x-x_0),
\label{p1Db}
\end{eqnarray}
\endnumparts
and $f(x_0,t)=D\partial_xp(0,t|x_0)$ the associated FPT density for a totally absorbing boundary. The first term on the right-hand side of equation (\ref{renewal}) represents all sample trajectories that have never hit the boundary at $x=0$ up to time $t$. This is independent of the state of the gate. The corresponding integral represents all trajectories that first reach the boundary at time $\tau$ with the gate closed, which occurs with probability $\Pi_{-1,k_0}(\tau)f(x_0,\tau)d\tau$, after which the particle instantaneously resets to position $x=\epsilon$. The particle can then return an arbitrary number of times before encountering an open gate and being absorbed. Indeed, iterating equation (\ref{renewal}) leads to the series solution
  \begin{eqnarray}
\fl  &\rho^{(\epsilon)}_{k|k_0}(x,t|x_0)\nonumber \\
 \fl &=\Pi_{kk_0}(t)p(x,t|x_0)+\int_0^t d\tau\,  \Pi_{k,-1}(t-\tau)p(x,t-\tau|\epsilon)\Pi_{-1,k_0}(\tau)  f(x_0,\tau) \nonumber  \\
 \fl &+\int_0^t d\tau  \int_0^{t-\tau}d\tau' \bigg [\Pi_{k,-1}(t-\tau-\tau')p(x,t-\tau-\tau' |\epsilon)\Pi_{-1,-1}(\tau+\tau') f(\epsilon,\tau')\nonumber \\
 \fl &\hspace{3cm} \times \Pi_{-1,k_0}(\tau)  f(x_0,\tau) \bigg ]+\ldots, 
  \label{itrenewal}
  \end{eqnarray}
 where the $n$th term in the series represents the contribution from sample path that hit the boundary $n-1$ times in the closed state before being absorbed on the $n$th visit, see Fig. \ref{fig3}. That is, the particle undergoes $n-1$ rounds of adsorption/desorption.

It is convenient to define the composite densities
 \begin{equation}
 \label{PiG2}
\fl  p_{k|k_0}(x,t|x_0)=\Pi_{kk_0}(t)p(x,t|x_0),\quad f_{k_0}(x_0,t)=\Pi_{-1,k_0}(t)f(x_0,t),
 \end{equation}
 with $\Pi_{kk_0}(t)$ given by equation (\ref{Pi}). Note that 
 \begin{equation}
 \sum_{k=\pm 1} p_{k|k_0}(x,t|x_0)=p(x,t|x_0),
 \end{equation}
 since $\sum_{k=\pm 1}\Pi_{kk_0}(t)=1$.
 Laplace transforming the renewal equation (\ref{renewal}) with respect to time $t$ and using the convolution theorem then gives
 \begin{eqnarray}
  \label{renewal2}
 \widetilde{\rho}^{(\epsilon)}_{k|k_0}(x,s|x_0) = \p_{k|k_0}(x,s|x_0)+ \widetilde{\rho}^{(\epsilon)}_{k|-1}(x,s|\epsilon)  \widetilde{f}_{k_0}(x_0,s) ,
 \end{eqnarray}
 where
 \begin{eqnarray}
\label{pk}
 \p_{k |k_0 }(x,s|x_0) 
&=\int_0^{\infty}\e^{-st}\Pi_{kk_0}(t) p(x,t|x_0)dt\nonumber \\
  &= \int_0^{\infty}\e^{-st}\bigg [\sigma_k+ (\delta_{k,k_0}-\sigma_k)\e^{-(\alpha+\beta) t }\bigg ]   p(x,t|x_0)dt \nonumber\\
 &=\sigma_k \p(x,s|x_0)+kk_0\sigma_{-k_0}\p(x,s+\alpha+\beta|x_0)  . 
\end{eqnarray}
Similarly,
 \begin{eqnarray}
 \label{fk}
 \widetilde{f}_{k_0}(x_0,s)&=\sigma_{-1}\f(x_0,s)-k_0\sigma_{-k_0} \pcb{\f(x_0,s+\alpha+\beta)}.
 \end{eqnarray}
 Laplace transforming equations (\ref{p1Da}) and (\ref{p1Db}) we see that $\p(x,s|x_0)=\G(x,s|x_0)$, where $G(x,s|x_0)$ is the Dirichlet Green's function (\ref{DG}), and
 \begin{eqnarray}
 \widetilde{f}(x_0,s)&=  {D}\partial_x \G(0,s|x_0)= \e^{-\sqrt{s/D} x_0} .
 \end{eqnarray}
Next, setting $x_0=\epsilon$, $k_0=-1$ in equation (\ref{renewal2}) and rearranging shows that
  \begin{equation}
 \widetilde{\rho}^{(\epsilon)}_{k|-1}(x,s|\epsilon) =\frac{ \p_{k|-1}(x,s|\epsilon)}{1-\widetilde{f}_{-1}(\epsilon,s)} .
 \end{equation}
  Substituting back into equations (\ref{renewal2}) yields the explicit solution
 \begin{eqnarray}
  \label{renewal3}
  \widetilde{\rho}^{(\epsilon)}_{k|k_0}(x,s|x_0) = \p_{k|k_0}(x,s|x_0)+\widetilde{f}_{k_0}(x_0,s)  \frac{ \p_{k|-1}(x,s|\epsilon) }{1-\widetilde{f}_{-1}(\epsilon,s)} .
 \end{eqnarray}
 Finally, taking the limit $\epsilon \rightarrow 0$ we have
  \begin{eqnarray}
  \label{alt2}
 \fl  \widetilde{\rho}_{k|k_0}(x,s|x_0)\equiv \lim_{\epsilon \rightarrow 0} \widetilde{\rho}^{(\epsilon)}_{k|k_0}(x,s|x_0) = \p_{k|k_0}(x,s|x_0)+\widetilde{f}_{k_0}(x_0,s) \lim_{\epsilon \rightarrow 0}\frac{ \p_{k|-1}(x,s|\epsilon) .}{1-\widetilde{f}_{-1}(\epsilon,s)}  .
 \end{eqnarray}
 Since $\p_{k|-1}(x,s|\epsilon)\rightarrow 0$ and $\widetilde{f}_{-1}(\epsilon,s)\rightarrow 1$ as $\epsilon \rightarrow 0$, the limit on the right-hand side of equation (\ref{alt2}) has to be evaluated using L'h\^opital's rule.
 \begin{eqnarray}
 \lim_{\epsilon \rightarrow 0}\frac{ \p_{k|-1}(x,s|\epsilon) }{1-\widetilde{f}_{-1}(\epsilon,s)}  =\frac{\sigma_k\e^{-\sqrt{s/D} x}-k\sigma_1 \e^{- \sqrt{s+2\alpha]/D} x}}{\sigma_{-1}\sqrt{sD}+\sigma_1\sqrt{[s+2\alpha]D}}.
 \end{eqnarray}
It can be checked that the solution (\ref{alt2}) for $k=\pm 1$ is equivalent to equation (\ref{sol}), which is the solution obtained directly from the forward Kolmogorov equation.

 We also note that summing both sides of equation (\ref{alt2}) with respect to $k=\pm 1$ and integrating with respect to $x\in [0,\infty)$ gives the following equation for the stochastically gated survival probability in Laplace space:
 \begin{eqnarray}
  \label{alt2Q}
  \widetilde{\mathfrak Q}_{k_0}(x_0,s)= \Q(x_0,s)+\widetilde{f}_{k_0}(x_0,s)  \lim_{\epsilon \rightarrow 0}\frac{ \Q(\epsilon, s)}{1-\widetilde{f}_{-1}(\epsilon,s)}  ,
 \end{eqnarray} 
 where (see equation (\ref{S}))
 \begin{eqnarray}
\widetilde{ \calQ}_{k_0}(x_0,s)&:=\int_0^{\infty}\sum_{k=\pm 1}\wrho_{k|k_0}(x,s|x_0) dx  ,
 \end{eqnarray} 
 and
\begin{eqnarray}
\Q(\epsilon,s)=\sum_{k=\pm 1} \int_0^{\infty} \p_{k|-1}(x,s|\epsilon)dx &  
 =\int_0^{\infty} p(x,s|\epsilon)dx
 \label{Qdef} 
 \end{eqnarray} 
 is the Laplace transformed survival probability with respect to adsorption.
 We have reversed the order of integration and the limit $\epsilon \rightarrow 0$ in equation (\ref{alt2Q}). Again the limit is evaluated using L'H\^opital's rule. In the case of the half-line, we have $\Q=\Q_0$ with
 \begin{eqnarray}
 \Q_0(x_0,s)&  
 :=\int_0^{\infty} \G(x,s|x_0)dx=\frac{1-\e^{-\sqrt{s/D}x_0}}{s} ,
 \label{Q0} \end{eqnarray} 
and
  \begin{eqnarray}
  \label{limQ}
 \lim_{\epsilon \rightarrow 0}\ \frac{ \Q_0(\epsilon, s)}{1-\widetilde{f}_{-1}(\epsilon,s)}  .  =\frac{ 1 }{\sigma_{-1}s+\sigma_1\sqrt{[s+\alpha+\beta]s}}.
 \end{eqnarray}

\subsubsection*{\bf Remarks}

\noindent (i) The renewal equation (\ref{renewal}) provides a general probabilistic formulation of a stochastically gated boundary. A major advantage compared to the forward Kolmogorov equation (\ref{1Da})--(\ref{1Dc}), is that the renewal equation explicitly separates the FPT problem for detecting the gated boundary from the mechanism for subsequently restarting the diffusion process. For example, the particle could reset to its initial position $x_0$ whenever it reflects off a closed gate, analogous to desorption-induced resetting \cite{Bressloff25b}. This means setting $\epsilon =x_0$ in equation (\ref{alt2Q}) rather than taking the limit $\epsilon \rightarrow 0$. In addition, the basic structure of the renewal equation still holds under various modifications of the diffusion process prior to each adsorption event. The only difference is in the construction of the densities $p(x,t|x_0)$ and $f(x_0,t)$. For example, the particle could reset to its initial position at a random sequence of times generated by a Poisson process (see section 3). This bulk resetting is distinct from desorption-induced resetting. Alternatively, the condition for terminating each round of bulk diffusion could be modified by setting $p=p^{(\kappa_0)}$ where $p^{(\kappa_0)}$ satisfies a Robin boundary condition with reactivity $\kappa_0$:
\numparts
\begin{eqnarray}
\label{1Dpkapa}
\fl \frac{\partial p^{(\kappa_0)}(x,t|x_0)}{\partial t}&=D\frac{\partial^2p^{(\kappa_0)}((x,t|x_0)}{\partial x^2},\quad 0<x<\infty,\\
\fl D\frac{\partial p^{(\kappa_0)}((0,t|x_0)}{\partial x}&=\kappa_0  p^{(\kappa_0)}(0,t|x_0),\quad p^{(\kappa_0)}(x,0|x_0)=\delta(x-x_0).
\label{1Dpkapb}
\end{eqnarray}
\endnumparts
In contrast to the case of a totally adsorbing boundary ($\kappa_0\rightarrow \infty$), if the particle is adsorbed when the gate is closed then the diffusive process can be restarted at $x=0$ without any shift in position. The renewal equation (\ref{renewal}) thus  becomes
 \begin{eqnarray}
\fl  \rho^{(\kappa_0)}_{k|k_0}(x,t|x_0)&=\Pi_{kk_0}(t)p^{(\kappa_0)}(x,t|x_0)+\int_0^t  \rho^{(\kappa_0)}_{k|-1}(x,t-\tau|0)\Pi_{-1,k_0}(\tau)  f^{(\kappa_0)}(x_0,\tau)d\tau ,\nonumber \\
\fl  \label{renkap}
 \end{eqnarray}
 with $f^{(\kappa_0)}(x_0,\tau)=\kappa_0 p^{(\kappa_0)}(x,\tau|x_0)$.
Again the renewal equation can be solved in Laplace space to give
\begin{eqnarray}
  \label{renkap2}
  \widetilde{\rho}^{(\kappa_0)}_{k|k_0}(x,s|x_0) = \p^{(\kappa_0)}_{k|k_0}(x,s|x_0)+\widetilde{f}^{(\kappa_0)}_{k_0}(x_0,s)  \frac{ \p_{k|-1}^{(\kappa_0)}(x,s|0) }{1-\widetilde{f}^{(\kappa_0)}_{-1}(0,s)} .
 \end{eqnarray}
 This also provides an alternative method for recovering the solution of the original forward Kolmogorov equation by taking the limit $\kappa_0\rightarrow \infty$ and using L'H\^opital's rule. 
\medskip

  \begin{figure}[t!]
  \centering
  \includegraphics[width=9cm]{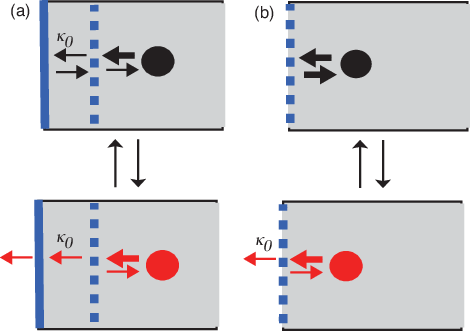}
  \caption{Schematic illustration comparing (a) partial adsorption prior to encountering the gate and (b) partial absorption through an open gate. Reduction of a thick arrow indicates partial reflection.}
  \label{fig4}
\end{figure}

 \noindent (iii) The renewal equation (\ref{renkap}) for finite $\kappa_0$ assumes that there is a necessary intermediate step (partial adsorption) prior to detecting the state of the gate, see Fig. \ref{fig4}(a).
 This differs from the classical treatment of stochastically gated chemical reactions developed by Szabo {\em et al.}
\cite{Szabo84,Spouge96}, see also Ref. \cite{Scher24}. These authors treated the closed gate as totally reflecting and the open gate as partially absorbing, see Fig. \ref{fig4}(b). In other words, $ \rho^{(\kappa_0)}_{-1|k_0}(x,t|x_0)$ and $ \rho^{(\kappa_0)}_{1|k_0}(x,t|x_0)$ satisfiied Neumann and Robin boundary conditions, respectively. Green's function methods were then used to construct an integral equation relating the propagator for $\kappa_0>0$ to the propagator for a totally reflecting boundary ($\kappa_0=0$). It is instructive to determine the corresponding boundary conditions satisfied by the solution of the renewal equation (\ref{renkap2}) when $\kappa_0$ is finite. First note that $\widetilde{f}^{(\kappa_0)}_{k_0}(x_0,s) =\kappa_0 \p^{(\kappa_0)}_{-1|k_0}(0,s|x_0)$. Second, the solution of equations (\ref{1Dpkapa}) and (\ref{1Dpkapb}) is \cite{Bressloff22}
 \begin{equation}
\label{EBM}
\fl \p(x,s|x_0)=\frac{1}{2\sqrt{sD}}\left (\e^{-\sqrt{s/D}|x-x_0|}+\frac{\sqrt{sD}-\kappa_0}{\sqrt{sD}+\kappa_0}\e^{-\sqrt{s/D}(x+x_0)}\right ),
\end{equation} 
which implies that
\begin{equation}
\p(0,s|x_0)=\frac{1}{\sqrt{sD}+\kappa_0}\e^{-\sqrt{s/D}x_0}.
\end{equation}
We also have a modified Robin boundary condition when $x_0=0$ \cite{Bressloff22}:
\begin{equation}
\left. D\frac{\partial \p(x,s|0)}{\partial x}\right |_{x=0}=\kappa_0 \p(0,s|0)-1.
\end{equation}
Combining these various result shows that
\begin{eqnarray}
\fl \left . D\frac{\partial  \widetilde{\rho}^{(\kappa_0)}_{k|k_0}(x,s|x_0)}{\partial x} \right |_{x=0}= \kappa_0 \p^{(\kappa_0)}_{k|k_0}(0,s|x_0)+\kappa_0 \p^{(\kappa_0)}_{-1|k_0}(0,s|x_0)  \frac{ \kappa_0\p_{k|-1}^{(\kappa_0)}(0,s|0) -1}{1-\kappa_0 \p^{(\kappa_0)}_{-1|-1}(0,s|0)  } .
 \end{eqnarray}
 It immediately follows that
 \begin{eqnarray}
\left . D\frac{\partial  \widetilde{\rho}^{(\kappa_0)}_{-1|k_0}(x,s|x_0)}{\partial x} \right |_{x=0}=0,
\end{eqnarray}
as expected. On the other hand, $\widetilde{\rho}^{(\kappa_0)}_{1|k_0}(x,s|x_0)$ does not satisfy a classical Robin boundary condition, reflecting the difference between the two schemes illustrated in Fig. \ref{fig4}.
 \medskip

 \noindent (iii) There is another interesting connection between stochastically gated and Robin boundary conditions, namely, the latter can be obtained from the former by taking
  the double limit $\alpha\rightarrow \infty$ and $\beta \rightarrow \infty$ under the additional constraint that the proportion of time that the gate spends in the closed state approaches unity. This was previously proven in the case of a macroscopic 
 model of diffusion by taking expectations of particle concentration with respect to realisations of the stochastic gate dynamics \cite{Lawley15a}. Here we provide a simpler single-particle derivation based on the solution of the Kolmogorov equation in Laplace space.
The basic idea is to introduce a fixed fundamental rate $\gamma_0$ and scale the transition rates $\alpha,\beta$ as
\begin{equation}
\label{scale}
\alpha = { (1-\eta)\gamma_0 }/{\delta},\quad \beta = {\eta \gamma_0 }/{\delta},
\end{equation}
with $0<\eta <1$ and $0<\delta \ll 1$. Ignoring short-time transients, we can take the Laplace variable $s$ to be bounded from above and $\delta$ to be sufficiently small so that $\alpha+\beta \gg s$. From equations (\ref{q}) and (\ref{cond}) we have
\begin{eqnarray}
\label{sol2}
 q(0,s)=A(s),\quad D\partial_xq(0,s)=\e^{-\sqrt{s/D}x_0}-\sqrt{sD} A(s),
 \end{eqnarray}
with
 \begin{eqnarray}
A(s)\sim  \frac{(1-\eta)\e^{-\sqrt{s/D}x_0}}{\eta\sqrt{(\alpha+\beta)D}}=\sqrt{\frac{\delta}{\gamma_0 D}}\frac{(1-\eta)\e^{-\sqrt{s/D}x_0}}{\eta}.
 \end{eqnarray}
 Hence,
 \begin{eqnarray}
 D\partial_xq(0,s)\sim \e^{-\sqrt{s/D}x_0}\sim \sqrt{\frac{\gamma_0 D}{\delta}}\frac{\eta}{1-\eta}q(0,s).
 \end{eqnarray}
A Robin boundary condition with effective permeability $\kappa_{\rm eff}$ is then obtained in the limit $\delta \rightarrow 0$ if 
\begin{equation}
\label{eta}
\eta\sim  \kappa_{\rm eff}\sqrt{ {\delta}/{\gamma_0 D}}.
\end{equation}
In other words, $\alpha = O(1/\delta)$ and $\beta=O(1/\sqrt{\delta})$ as $\delta \rightarrow 0$.

 \section{Renewal equation for 1D FPT problems}
 
 In this section we consider two applications of the renewal formulation to FPT problems for Brownian motion in 1D. In order to simplify the algebra we assume $\alpha=\beta$ throughout so that $\sigma_k=1/2$ for $k=\pm 1$.
 
 \subsection{Brownian particle with stochastic resetting on the half-line} 
 
Let us first return to the example of a Brownian particle on a half-line with a stochastically gated boundary at $x=0$. However, we now assume that when the particle is diffusing in the bulk domain $(0,\infty)$ it instantaneously resets to a fixed reset position $x_r$ at a random sequence of times generated from a Poisson process with constant rate $r$ \cite{Evans11a,Evans11b}. (We suppress the explicit dependence on $x_r$. At the end of the calculation we will set $x_r=x_0$.) The first renewal equation (\ref{renewal}) still holds except that the forward Kolmogorov equation for $p$ becomes
\numparts
\begin{eqnarray}
\label{reseta}
 \fl &\frac{\partial p(x,t|x_0)}{\partial t}=D\frac{\partial^2 p(x,t|x_0)}{\partial x^2}-rp(x,t|x_0)+rQ(x_0,t)\delta(x-x_r),\\
\fl & p(0,t|x_0)=0,\quad p(x,0|x_0)=\delta(x-x_0),
 \label{resetb}
\end{eqnarray}
\endnumparts
where $Q(x_0,t)$ is the survival probability 
\begin{equation}
\label{Sr}
 Q(x_0,t)=\int_0^{\infty} p(x,t|x_0)dx.
\end{equation}
Equations (\ref{reseta}) and (\ref{resetb}) can either be solved directly using Laplace transforms or by using renewal theory \cite{Evans11a,Evans11b}. Here we use the former approach. Laplace transforming equations (\ref{reseta}) and (\ref{resetb}) gives
\begin{eqnarray}
\label{resetLT}
\fl  &D\frac{\partial^2\p(x,s|x_0)}{\partial x^2} -(r+s)\p(x,s|x_0) 
  =-\delta(x-x_0)-r \widetilde{Q}(x_0,s)\delta(x-x_r),\quad x>0,
 \end{eqnarray}
 with $\p(0,s|x_0)=0$. It immediately follows that
  \begin{eqnarray}
   \p(x, s|x_0) &= \G(x,r+s|x_0)+ r\widetilde{Q}(x_0,s)  \G(x,r+s|x_r),
\label{p1Dr}
\end{eqnarray}
with $\G(x,s|x_0)$ given by equation (\ref{DG}).
 Finally, Laplace transforming equation (\ref{Sr}) and using (\ref{p1Dr}) shows that
 \begin{eqnarray}
\fl   \widetilde{Q}(x_0,s)&=\int_0^{\infty}\p(x,s|x_0)dx = \widetilde{Q}_0(x_0,r+s)+r\widetilde{ Q}(x_0,s)\widetilde{Q}_0(x_r,r+s),
\label{SSr}
 \end{eqnarray}
 where $\widetilde{Q}_0(x_0,s)$ is the survival probability without resetting, see equation (\ref{Q0}).
 Rearranging equation (\ref{SSr}) then gives the well-known results \cite{Evans11a,Evans11b}
  \numparts
   \begin{eqnarray}
     \label{Qrena}
\fl \widetilde{Q}(x_0,s)&=\frac{\widetilde{Q}_0(x_0,r+s)}{1-r\widetilde{Q}_0(x_r,r+s)},\\
\fl     \widetilde{f}(x_0,s)&\equiv  \frac{1-(r+s)\widetilde{Q}_0(x_0,r+s) }{1-r\widetilde{Q}_0(x_r,r+s) } 
     +\frac{r\bigg[\widetilde{Q}_0(x_0,r+s)  -\widetilde{Q}_0(x_r,r+s)  \bigg ]}{1-r\widetilde{Q}_0(x_r,r+s)}.
       \label{Qrenb}
\end{eqnarray}
\endnumparts

In order to take the limit $\epsilon \rightarrow 0$ in equation (\ref{alt2Q}) we need to determine the leading order contributions to $\Q(\epsilon, s)$ and $\f(\epsilon,s)$. Hence, setting $x_0=\epsilon$ in equations (\ref{Qrena}) and  (\ref{Qrenb}), we have
\numparts
 \begin{eqnarray}
 \label{resetFP2a}
   \Q(\epsilon, s) &=\frac{\epsilon/\sqrt{[r+s]D} }{1-r\widetilde{Q}_0(x_r,r+s)} +O(\epsilon^2),
   \   \end{eqnarray}
   and
   \begin{eqnarray}
  \fl  \widetilde{f}(\epsilon,s)&\equiv  \frac{1-\epsilon\sqrt{[r+s]/D}}{1-r\widetilde{Q}_0(x_r,r+s) } 
     +\frac{r\bigg[\epsilon/\sqrt{[r+s]D} -\widetilde{Q}_0(x_r,r+s)  \bigg ]}{1-r\widetilde{Q}_0(x_r,r+s)}+O(\epsilon^2)\nonumber \\
 \fl     &=1-\frac{\epsilon }{1-r\widetilde{Q}_0(x_r,r+s)}\frac{s}{\sqrt{[r+s]D}}+O(\epsilon^2).
   \label{resetFPb}
\end{eqnarray}
\endnumparts
Combining with equation (\ref{fk}) then yields
  \begin{eqnarray}
\fl & \lim_{\epsilon \rightarrow 0}\ \frac{ \Q(\epsilon, s)}{1-\widetilde{f}_{-1}(\epsilon,s)}  \\
\fl & \quad =\frac{\displaystyle \frac{2/\sqrt{[r+s]D} }{1-r\widetilde{Q}_0(x_r,r+s)}}{\displaystyle \frac{1}{1-r\widetilde{Q}_0(x_r,r+s)}\frac{s}{\sqrt{(r+s)D}}+\frac{1}{1-r\widetilde{Q}_0(x_r,r+s+2\alpha)}\frac{(s+2\alpha)}{\sqrt{(r+s+2\alpha)D}}}. \nonumber
  \end{eqnarray}
Now taking the limit $s\rightarrow 0$ in equation (\ref{alt2Q}) and setting $x_r=x_0$ yields the MFPT
\begin{eqnarray}
\fl \calT_{k_0}(x_0)&=\frac{ \widetilde{Q}_0(x_0,r)}{1-r\widetilde{Q}_0(x_0,r)} +\f_{k_0}(x_0,0)\frac{\e^{\sqrt{r/D}x_0}}{\sqrt{r}}\frac{\sqrt{r+s+2\alpha}}{\alpha}\bigg [1-r\widetilde{Q}_0(x_0,r+2\alpha)\bigg ],\nonumber \\
\fl\end{eqnarray}
and
\begin{equation}
\f_{k_0}(x_0,0)=\frac{1}{2} \left (1 -k_0 \frac{1-(r+2\alpha )\widetilde{Q}_0(x_0,r+2\alpha ) }{1-r\widetilde{Q}_0(x_0,r+2\alpha ) }\right ).
\end{equation}
Substituting for $\f_{k_0}(x_0,0$ and performing some algebra yields the final result
\begin{eqnarray}
\label{Treset}
\fl \calT_{k_0}(x_0)&=\frac{ \widetilde{Q}_0(x_0,r)}{1-r\widetilde{Q}_0(x_0,r)} \nonumber \\
\fl &\quad +\frac{\e^{\sqrt{r/D}x_0}}{\sqrt{r}}\frac{\sqrt{r+2\alpha}}{2\alpha}\bigg [1-k_0+[k_0(r+2\alpha)-r]\widetilde{Q}_0(x_0,r+2\alpha)\bigg ].\end{eqnarray}
It can be checked that equation (\ref{Treset}) recovers the MFPT for $k_0=\pm 1$ derived in Ref. \cite{Boyer21b} by directly solving the forward Kolmogorov equation for $\rho_{k|k_0}$ in the presence of resetting. Also note that in the slow switching limit, we have
\begin{equation}
\fl \lim_{\alpha \rightarrow 0} \calT_{-1}(x_0)=\infty,\quad \lim_{\alpha \rightarrow 0} \calT_{1}(x_0)=\frac{ 2\widetilde{Q}_0(x_0,r)}{1-r\widetilde{Q}_0(x_0,r)} =2\frac{\e^{\sqrt{r/D}x_0}-1}{r}.
\end{equation}
The first result is expected: the gate is closed almost surely when $k_0=-1$ and $\alpha \rightarrow 0$ so the MFPT diverges  The second result is less intuitive: although the gate is open almost surely when $k_0=1$, the MFPT is double the result obtained for a totally absorbing boundary (no gate). This is a consequence of a rare event (closed gate) resulting in a large contribution to the MFPT such that the combination of the two factors is finite in the limit $\alpha \rightarrow 0$. (This limit is distinct from the fast switching limit used to obtain Robin boundary condition, see remark at end of section 2.)

\begin{figure}[t!]
\centering
\includegraphics[width=13cm]{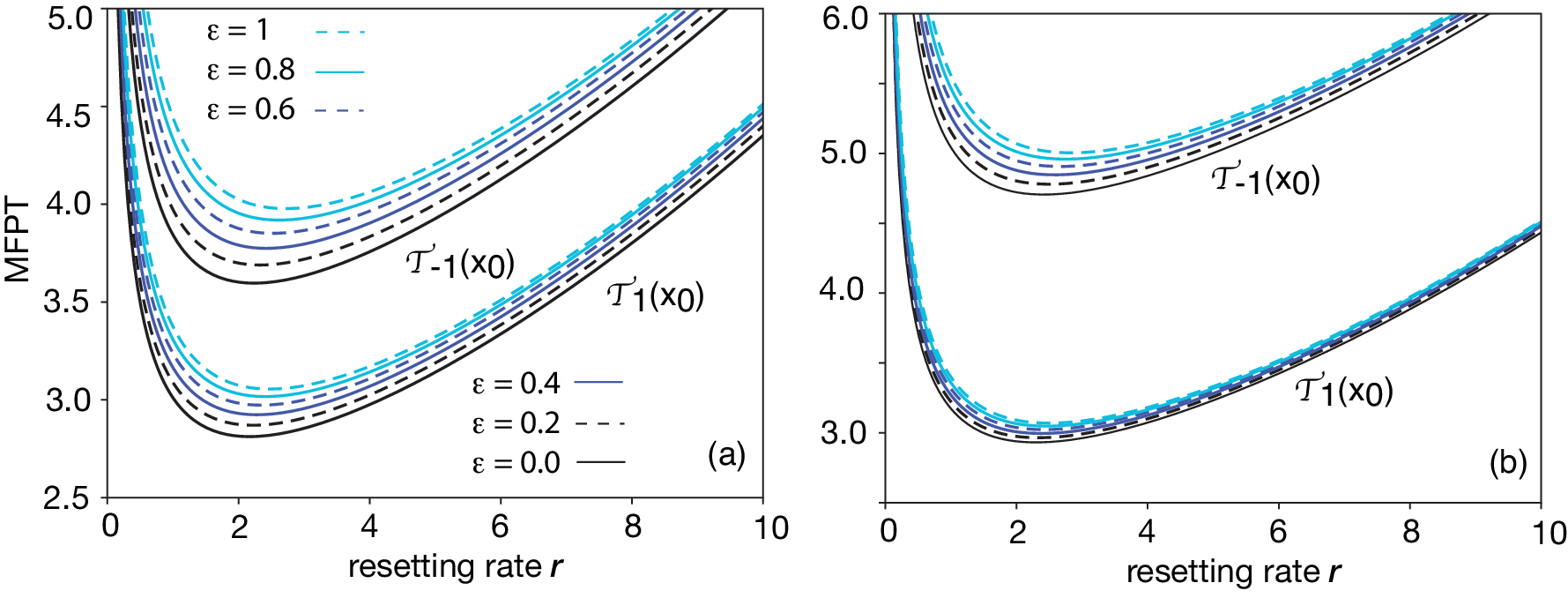} 
\caption{Plots of the MFPTs $\calT_{\pm 1}(x_0)$ as a function of the resetting rate $r$ for various values of $\epsilon$: (a) $\alpha =1$ and (b) $\alpha =0.5$. Other parameter values are $D=1$ and $x_0=1$.}
\label{fig5}
\end{figure}

One useful feature of the renewal formulation is that we can straightforwardly generalise the model by, for example, keeping $\epsilon >0$ fixed 
rather than taking the limit $\epsilon \rightarrow 0$ in equation (\ref{alt2Q}). (In Ref. \cite{Bressloff25b} we present an argument for a particle resetting to its initial position ($\epsilon =x_0$) when leaving a boundary without being absorbed, based on a foraging animal returning to its homebase after failing to capture resources.) Therefore, consider 
 \begin{eqnarray}
  \label{alt2Qreset}
  \widetilde{\mathfrak Q}_{k_0}(x_0,s)= \Q(x_0,s)+\widetilde{f}_{k_0}(x_0,s) \frac{ \Q(\epsilon, s)}{1-\widetilde{f}_{-1}(\epsilon,s)}  ,\quad \epsilon >0,
 \end{eqnarray} 
 with $\widetilde{f}_{k_0}(x_0,s) $ defined by equations (\ref{fk}) and (\ref{Qrenb}). In Fig. \ref{fig5} we show example plots of the corresponding MFPT $\calT_{k_0}(x_0)= \lim_{s\rightarrow 0}\widetilde{\mathfrak Q}_{k_0}(x_0,s)$ as a function of the resetting rate $r$ for different values of $\epsilon$ and $\alpha$. It can be seen that the solutions converge in the limit $\epsilon \rightarrow 0$. A number of other features can be observed. First, each curve has the characteristic unimodal shape indicating the existence of an optimal resetting rate for minimising the MFPT. Second, $\calT_1(x_0)<\calT_{-1}(x_0)$ for the same set of parameter values, that is, the MFPT is reduced if the gate is initially open. Third, reducing the switching rate $\alpha$ increases the gap between $\calT_1(x_0)$ and $\calT_{-1}(x_0)$ with $\lim_{\alpha \rightarrow 0}\calT_{-1}(x_0)=\infty$. 
  
 Finally, note that we have assumed that only the position of the particle resets whereas the state of the gate is unaffected. If the state of the gate also resets then it is possible to write down a renewal equation that relates the propagator $\rho_{k|k_0}$ with resetting and stochastic gating to the corresponding propagating with stochastic gating alone \cite{Bressloff20}. The latter was calculated in section 2.
 
 \subsection{Brownian particle in a finite interval}
 
 As our second example, consider a Brownian particle in the finite interval $[0,L]$ with a stochastically gated boundary at both ends. Following Refs. \cite{PCB15a,PCB15b,Bressloff16a}, we have to specify  whether the particle or the gates physically switch states. If the particle switches between a reactive and a non-reactive state, then either both gates are open or both gates are closed. That is, the two ends are totally correlated. On the other hand, if the individual gates independently switch between open and closed states then the two ends are completely uncorrelated. (This distinction also has major implications at the multi-particle level, even for a single gate. In particular, if the gate switches state, then all of the particles are subject to a common switching environment, which induces statistical correlations between the particles in spite of the fact that they are non-interacting \cite{PCB15a}. On the other hand, statistical correlations are absent in the case of independently switching, non-interacting particles.)
 
 \subsubsection{Correlated gates} Let us first consider the totally correlated case with $\Pi_{kk_0}(t)$ specifying the probability that the particle is in the state $k$ at time $t$, given that it was initially in state $k_0$.
  Using the casual structure of the renewal formulation, it is straightforward to write down the analogue of the first renewal equation (\ref{renewal}): 
  \begin{eqnarray}
 \fl \rho^{(\epsilon)}_{k|k_0}(x,t|x_0)&=\Pi_{kk_0}(t)p(x,t|x_0)-\int_0^t  \rho^{(\epsilon)}_{k|-1}(x,t-\tau|\epsilon)\Pi_{-1,k_0}(\tau)J(0,\tau|x_0)d\tau \nonumber \\
\fl  &\quad +\int_0^t  \rho^{(\epsilon)}_{k|-1}(x,t-\tau|L-\epsilon)\Pi_{-1,k_0}(\tau) J(L,\tau|x_0)d\tau.
 \label{ren2gate}
 \end{eqnarray}
Here $p(x,t|x_0)$ is the propagator for totally adsorbing boundaries at $x=0,L$:
\numparts
\begin{eqnarray}
\label{1Dp2gatea}
&  \frac{\partial p(x,t|x_0)}{\partial t}=D\frac{\partial^2p(x,t|x_0)}{\partial x^2},\quad 0<x<L,\\
&p(0,t|x_0)=0=p(L,t|x_0),\quad p(x,0|x_0)=\delta(x-x_0),
\label{1Dp2gateb}
\end{eqnarray}
\endnumparts
and $J(x,t|x_0)=-D\partial_xp(x,t|x_0)$. 
The first integral on the right-hand side represents all sample paths that (i) reach $x=0$ for the first time prior to ever reaching $x=L$ and (ii) the particle is in a non-reactive state (closed gate). The second integral corresponds to the complementary scenario in which the particle reaches $x=L$ first. As in the case of the half-line, it is convenient to define the composite densities
 \begin{equation}
 \label{PiG2a}
\fl  p_{k|k_0}(x,t|x_0)=\Pi_{kk_0}(t)p(x,t|x_0),\quad J_{k_0}(aL,t|x_0)=\Pi_{-1,k_0}(t)J(aL,t|x_0),\  a=0,1.
 \end{equation}
 Note that the exit flux $J(L, t|x_0)$ at $x=L$ determines the splitting probability $\pi_L(x_0)$ and conditional FPT density  $f_L(x_0,t)$ for adsorption at $x=L$ according to
 \begin{equation}
\pi_L(x_0)=\int_0^{\infty} J(L,t|x_0)dt,\quad f_L(x_0,t)=\frac{J(L,t|x_0)}{\pi_L(0)}.
\end{equation}
The corresponding quantities at $x=0$ are
\begin{equation}
\pi_0(x_0)=-\int_0^{\infty} J(0,t|x_0)dt,\quad f_0(x_0,t)=-\frac{J(0,t|x_0)}{\pi_0}.
\end{equation}

Laplace transforming equation (\ref{ren2gate}) we have
\begin{eqnarray}
  \widetilde{\rho}^{(\epsilon)}_{k|k_0}(x,s|x_0) &= \p_{k|k_0}(x,s|x_0)- \widetilde{\rho}^{(\epsilon)}_{k|-1}(x,s|\epsilon)\J_{k_0}(0,s|x_0) \nonumber \\
 & + \widetilde{\rho}^{(\epsilon)}_{k|-1}(x,s|L-\epsilon) \J_{k_0}(L,s|x_0) ,  
  \label{ren2gate2}
\fl\end{eqnarray}
 where $\p_{k|k_0}(x,s|x_0)$ is given by equation (\ref{pk}) and
 \begin{eqnarray}
 \label{fk2gate}
 \widetilde{J}_{k_0}(x,s|x_0)&= \frac{1}{2}\bigg (\J(x,s|x_0)-k_0\J(x,s+2\alpha|x_0)\bigg ).
 \end{eqnarray}
In order to eliminate the terms $\widetilde{\rho}^{(\epsilon)}_{k|-1}(x,s|\epsilon) $ and $\widetilde{\rho}^{(\epsilon)}_{k|-1}(x,s|\epsilon) $ on the right-hand side of equation (\ref{ren2gate2}) we set $k_0=-1$ and either $x_0=\epsilon$ or $x_0=L-\epsilon$. This leads to the pair of simultaneous equations
\numparts
\begin{eqnarray}
 \widetilde{\rho}^{(\epsilon)}_{k|-1}(x,s|\epsilon) &= \p_{k|-1}(x,s|\epsilon)- \widetilde{\rho}^{(\epsilon)}_{k|-1}(x,s|\epsilon) \widetilde{J}_{-1}(0,s|\epsilon) \nonumber \\
 &\quad + \widetilde{\rho}^{(\epsilon)}_{k|-1}(x,s|L-\epsilon) \widetilde{J}_{-1}(L,s|\epsilon) ,\\
 \widetilde{\rho}^{(\epsilon)}_{k|-1}(x,s|L-\epsilon) &= \p_{k|-1}(x,s|L-\epsilon)- \widetilde{\rho}^{(\epsilon)}_{k|-1}(x,s|\epsilon) \widetilde{J}_{-1}(0,s|L-\epsilon)
\nonumber \\
&\quad + \widetilde{\rho}^{(\epsilon)}_{k|-1}(x,s|L-\epsilon)\widetilde{J}_{-1}(L,s|L-\epsilon).
 \end{eqnarray}
 \endnumparts
 We thus have the formal solution
 \begin{eqnarray}
 \label{solA}
\fl  \left (\begin{array}{c} \widetilde{\rho}^{(\epsilon)}_{k|-1}(x,s|\epsilon)\\ \widetilde{\rho}^{(\epsilon)}_{k|-1}(x,s|L-\epsilon)\end{array}\right )=\left [
 {\bf I}-{\bf M}(\epsilon,s) \right ]^{-1}\left (\begin{array}{c} \widetilde{p} _{k|-1}(x,s|\epsilon)\\ \widetilde{p} _{k|-1}(x,s|L-\epsilon)\end{array}\right ) ,
 \end{eqnarray}
 where
 \begin{eqnarray}
 \label{M}
 {\bf M}(\epsilon, s):=\left (\begin{array}{cc} -\widetilde{J}_{-1}(0,s|\epsilon) & \widetilde{J}_{-1}(L,s|\epsilon)\\ - \widetilde{J}_{-1}(0,s|L-\epsilon)
 & \widetilde{J}_{-1}(L,s|L-\epsilon) \end{array} \right ) .
  \end{eqnarray}
  Substituting the solution (\ref{solA}) into equation (\ref{ren2gate2}) and taking the limit $\epsilon \rightarrow 0$ gives
  \begin{eqnarray}
  \label{ren2gate3}
 \fl \widetilde{\rho}_{k|k_0}(x,s|x_0) &= \p_{k|k_0}(x,s|x_0)\\
\fl  &\quad -
 \sum_{a,b=0,1}(-1)^a \J_{k_0}(aL,s|x_0)\lim_{\epsilon\rightarrow 0}\left [
 {\bf I}-{\bf M}(\epsilon,s) \right ]_{ab}^{-1}\widetilde{p} _{k|-1}(x,s| |bL-\epsilon|).\nonumber 
 \end{eqnarray}
 Finally, integrating with respect to $x$ and summing with respect to $k$ we obtain the solution for the Laplace transformed survival probability according to
  \begin{eqnarray}
\fl  \widetilde{\calQ}_{k_0}(x_0,s) &=\Q(x_0,s)-\lim_{\epsilon \rightarrow 0}
 \sum_{a,b=0,1} (-1)^a\J_{k_0}(aL,s|x_0)\left [
 {\bf I}-{\bf M}(\epsilon,s) \right ]_{ab}^{-1}\Q(|bL-\epsilon|,s).\nonumber \\
 \fl  \label{ren2gateQ}
 \end{eqnarray}
 Here
 \numparts
 \begin{eqnarray}
 \label{Qdef2gate} 
 \calQ_{k_0}(x_0,s)&:=\int_0^{L}\sum_{k=\pm 1}\wrho_{k|k_0}(x,s|x_0) dx  ,\\
  \Q(x_0,s)  
 &:=\int_0^L \p(x,s|x_0)dx=\frac{1-\J(L,s|x_0)+\J(0,s|x_0)}{s}.
 \end{eqnarray} 
 \endnumparts
Equation (\ref{ren2gateQ}) is the 2-gate analogue of equation (\ref{alt2Q}).

 The propagator $\p(x,s|x_0)$ is equal to the Dirichlet Green's function of the modified Helmholtz equation on $[0,L]$ and has the classical expression
  \begin{eqnarray}
\label{G1D}
 G(x,s|x_0)&=\frac{\sinh \sqrt{s/D}x\, \sinh \sqrt{s/D}(L-x_0) }{\sqrt{sD}\sinh \sqrt{s/D}L}\Theta(x_0-x)\\
 &\quad + \frac{\sinh \sqrt{s/D} x_0\, \sinh \sqrt{s/D}(L-x) }{\sqrt{sD}\sinh \sqrt{s/D}L}\Theta(x-x_0),
\end{eqnarray}
 where $\Theta(x)$ is a Heaviside function. It follows that
 \begin{eqnarray}
 \fl \J(0,s|x_0)&=-\frac{ \sinh \sqrt{s/D}(L-x_0) }{ \sinh \sqrt{s/D}L},\quad \J(L,s|x_0)=\frac{ \sinh \sqrt{s/D} x_0}{ \sinh \sqrt{s/D}L}.\end{eqnarray}
 Substituting into the matrix (\ref{M}) and expanding to $O(\epsilon)$ gives
  \begin{eqnarray}
 \label{M2}
 {\bf M}(\epsilon, s)={\bf I}-\epsilon {\bm \Lambda}(s)+O(\epsilon^2),
  \end{eqnarray}
  with ${\bm \Lambda}(s)={\bf m}(s)+ {\bf m}(s+2\alpha)$ and
   \begin{eqnarray}
 \label{M3}
 {\bf m}(s) = \frac{ \sqrt{{s}/{D}}}{2\sinh\sqrt{s/D}L}\left (\begin{array}{cc} \mbox{cosh}\sqrt{s/D}L& -1\\ -1
 &  \mbox{cosh}\sqrt{s/D}L\end{array} \right ) .
  \end{eqnarray}
  Since, by symmetry,
  \begin{equation}
 \fl  \Q(L-\epsilon,s)=\Q(\epsilon,s)=\epsilon \q(s)+O(\epsilon^2),\quad \q(s)\equiv \frac{\cosh\sqrt{s/D}L-1}{\sqrt{sD}\sinh\sqrt{s/D}L},
  \end{equation} 
   it follows that
   \begin{eqnarray}
  \label{ren2gate4}
\lim_{\epsilon\rightarrow 0}\sum_{b=0,1}\left [
 {\bf I}-{\bf M}(\epsilon,s) \right ]_{ab}^{-1}\Q(|bL-\epsilon|,s)
=\q(s)\sum_{b=0,1}{\bm \Lambda}(s)_{ab}^{-1}.
 \end{eqnarray}

 Finally, substitute equation (\ref{ren2gate4}) back into equation (\ref{ren2gateQ}) and take the limit $s\rightarrow 0$
using
\numparts
 \begin{eqnarray}
\fl \lim_{s\rightarrow 0}\q(s)&= \frac{L}{2D} ,\\
\fl  \lim_{s\rightarrow 0}{\bm \Lambda}(s)^{-1}&= 2L\left (\begin{array}{cc} A(\alpha) & -B(\alpha) \\ \\ -B(\alpha)
 &  A(\alpha)\end{array} \right )^{-1} =\frac{2L}{A(\alpha)^2-B(\alpha)^2}\left (\begin{array}{cc} A(\alpha) &B(\alpha) \\ \\ B(\alpha)
&  A(\alpha) \end{array} \right ),\nonumber \\
\fl
\end{eqnarray} 
\endnumparts
with
\begin{equation}
A(\alpha)=1+ \frac{  \sqrt{2\alpha/D}L}{\tanh \sqrt{2\alpha/D}L},\quad B(\alpha)=1+\frac{  \sqrt{2\alpha/D}L}{\sinh\sqrt{2\alpha/D}L}.
\end{equation}
Since $\lim_{s\rightarrow 0}\sum_{b=0,1}{\bm \Lambda}(s)_{ab}^{-1}=2L/(A(\alpha)-B(\alpha))$ for $a=0,1$, we obtain the MFPT
 \begin{eqnarray}
\fl  \calT_{k_0}(x_0)&=T(x_0)+\frac{L^2}{D}\frac{\sinh \sqrt{2\alpha/D}L}{  \sqrt{2\alpha/D}L\bigg [\cosh \sqrt{2\alpha/D}L-1\bigg ]}\bigg [\J_{k_0}(L,0|x_0)-\J_{k_0}(0,0|x_0)\bigg ].\nonumber 
 \end{eqnarray}
Here $T(x_0)$ is the classical MFPT for a Brownian particle in $[0,L]$ with totally adsorbing boundaries at both ends,
\begin{eqnarray}
T(x_0)&=\lim_{s\rightarrow 0}\Q(x_0,s)=\lim_{s\rightarrow 0} \frac{1-\J(L,s|x_0)+\J(0,s|x_0)}{s}\nonumber \\
&=-\partial_s\J(L,0|x_0)+\partial_s\J(0,0|x_0)=\frac{x_0(L-x_0)}{2D}.
\end{eqnarray}
 
 \subsubsection{Uncorrelated gates} The case of uncorrelated gates is more complicated as we now have to keep track of two independent 2-state Markov chains. Let $k=\pm 1$ denote the state of the gate at $x=0$ with switching rate $\alpha$ and $\bar k$ denote the state of the gate at $x=L$ with switching rate $\bar \alpha$. The renewal equation takes the form
  \begin{eqnarray}
\fl \rho^{(\epsilon)}_{k\bar k|k_0\bar k_0}(x,t|x_0)&=\Pi_{kk_0}(t)\overline{\Pi}_{\bar k \bar k_0}(t)p(x,t|x_0)\nonumber \\
 \fl &\quad -\int_0^t  \rho^{(\epsilon)}_{k\bar k |-1,\bar k_0}(x,t-\tau|\epsilon)\Pi_{-1k_0}(\tau)  J(0,\tau|x_0)d\tau \nonumber \\
 \fl &\quad +\int_0^t  \rho^{(\epsilon)}_{k\bar k|k_0 ,-1}(x,t-\tau|\epsilon) \overline{\Pi}_{-1| \bar k_0}(\tau)  J(L,\tau|x_0)d\tau.
 \label{rencorr}
 \end{eqnarray}
 Note that if the particle reaches the left-hand gate first, then what happens next is independent of the state of the right-hand gate and vice versa. The transition matrix $\overline{\bm \Pi}(t)$ is given by equation (\ref{Pi}) with $\alpha,\beta \rightarrow \bar \alpha$. Define
 \numparts
 \label{PiG3}
 \begin{eqnarray}
\fl & p_{k\bar k|k_0\bar k_0}(x,t|x_0)=\Pi_{kk_0}(t)\overline{\Pi}_{\bar k \bar k_0}(t)p(x,t|x_0),\\ 
\fl & J_{k_0}(0,t|x_0)=\Pi_{-1,k_0}(t)J(0,t|x_0),\quad J_{\bar k_0}(L,t|x_0)=\overline{\Pi}_{-1,\bar k_0}(t)J(L,t|x_0).
 \end{eqnarray}
 \endnumparts
 Laplace transforming equation (\ref{rencorr}) we have
\begin{eqnarray}
 \widetilde{\rho}^{(\epsilon)}_{k\bar k|k_0 \bar k_0}(x,s|x_0) &= \p_{k\bar k|k_0\bar k_0}(x,s|x_0)- \widetilde{\rho}^{(\epsilon)}_{k\bar k|-1,\bar k_0}(x,s|\epsilon)\J_{k_0}(0,s|x_0) \nonumber \\
 &\quad + \widetilde{\rho}^{(\epsilon)}_{k\bar k|k_0,-1}(x,s|L-\epsilon) \J_{\bar k_0}(L,s|x_0) .
  \label{rencorr2}
  \end{eqnarray}
 Setting $k_0=-1$ and $x_0=\epsilon$ or $\bar k_0=-1$ and $x_0=L-\epsilon$ leads to the four simultaneous equations (for fixed $k,\bar k$)
\numparts
\begin{eqnarray}
\fl  \widetilde{\rho}^{(\epsilon)}_{k\bar k|-1,\bar k_0}(x,s|\epsilon) &= \p_{k \bar k|-1,k_0}(x,s|\epsilon)- \widetilde{\rho}^{(\epsilon)}_{k\bar k|-1,\bar k_0}(x,s|\epsilon) \widetilde{J}_{-1}(0,s|\epsilon) \nonumber \\
 \fl &\quad + \widetilde{\rho}^{(\epsilon)}_{k\bar k|-1,k_0}(x,s|L-\epsilon) \widetilde{J}_{\bar k_0}(L,s|\epsilon) , \quad \bar k_0=\pm 1,\\
\fl \widetilde{\rho}^{(\epsilon)}_{k\bar k|k_0,-1}(x,s|L-\epsilon) &= \p_{k \bar k|k_0,-1}(x,s|L-\epsilon)- \widetilde{\rho}^{(\epsilon)}_{k\bar k|k_0,-1}(x,s|\epsilon) \widetilde{J}_{k_0}(0,s|L-\epsilon)
\nonumber \\
\fl &\quad + \widetilde{\rho}^{(\epsilon)}_{k\bar k|k_0,-1}(x,s|L-\epsilon)\widetilde{J}_{-1}(L,s|L-\epsilon),\quad k_0=\pm 1 .
 \end{eqnarray}
 \endnumparts
 There are thus two major differences from the case of uncorrelated gates. First, the matrix ${\bf M}(\epsilon,s)$ defined in equation (\ref{M}) becomes four-dimensional. Second, equation (\ref{pk}) becomes
 \begin{eqnarray}
\fl&\p_{k\bar k|k_0\bar k_0}(x,s|x_0)\nonumber \\
\fl&=\int_0^{\infty}\e^{-st}\Pi_{kk_0}\overline{\Pi}_{\bar k \bar k_0}(t)p(x,t|x_0)dt\nonumber \\
\fl&= \int_0^{\infty}\e^{-st}\bigg [\frac{1}{2}+ (\delta_{k,k_0}-1/2)\e^{-2\alpha t }\bigg ]  \bigg [\frac{1}{2}+ (\delta_{\bar k,\bar k_0}-1/2)\e^{-2\alpha t }\bigg ]p(x,t|x_0)dt \nonumber \\
\fl &=\frac{1}{4}\bigg (\p(x,s|x_0)+kk_0\p(x,s+2\alpha|x_0)+\bar k \bar k_0\p(x,s+2\bar \alpha|x_0)\nonumber \\
\fl &\hspace{3cm}+kk_0\bar k \bar k_0 \p(x,s+2(\bar \alpha+\alpha)|x_0)\bigg ). 
\label{pkk}
\end{eqnarray}
We conclude that the algebra becomes considerably more involved as the number of targets increases. Nevertheless, the renewal formulation provides a systematic procedure for constructing solutions 
that could be implemented algorithmically.

 \section{Higher-dimensional renewal equation}
In order to construct a higher-dimensional renewal equation, we consider a simplified version of the configuration shown in Fg. \ref{fig1}(a) where the boundary $\partial \Omega \subset \R^d$ is a stochastically gated absorbing surface and there is no interior target $\calU$. Again we set $\alpha=\beta$ for simplicity. Suppose that whenever the particle reflects off a point $\y \in \partial \Omega$ when the gate is closed, it is infinitesimally shifted according to 
\begin{equation}
\y\rightarrow \a_{\epsilon}(\y)=\y-\epsilon \n(\y), 
\end{equation}
where $\n(\y)$ is the unit normal directed towards the exterior of $\Omega$. (Shifts tangential to the surface do not contribute to the limit $\epsilon \rightarrow 0$ because of the Dirichlet boundary condition.) The higher-dimensional version of the renewal equation (\ref{renewal}) takes the form
\begin{eqnarray}
 \fl  \rho^{(\epsilon)}_{k|k_0}(\x,t|\x_0) 
  &=\Pi_{k|k_0}(t)p(\x,t|\x_0)\nonumber \\
  \fl &\quad +\int_{\partial \Omega}d\y\int_0^td\tau \,   \rho^{(\epsilon)}_{k|-1}(\x,t-\tau|\a_{\epsilon}(\y))\Pi_{-1|k_0}(\tau) J(\y,\tau|\x_0). 
  \label{2Dren}
 \end{eqnarray}
 Here $p(\x,t|\x_0)$ is the propagator for a totally absorbing boundary $\partial \calU$:
 \begin{eqnarray}
 \fl \frac{\partial p(\x,t|\x_0)}{\partial t} &=D{\bm \nabla}^2 p(\x,t|\x_0),\quad \x \in \Omega ,\qquad
 p(\y,t|\x_0)=0 \quad \y \in \partial  \Omega .
 \end{eqnarray}
The term $J(\y,t|\x_0)=-D\left . {\bm \nabla}p(\x,t|\x_0) \right|_{\x=\y}\cdot \n(\y)$ is the corresponding adsorption probability flux into the target at $\y\in \partial \Omega$.
 Introduce the composite functions
 \begin{equation}
 \label{PiG2D}
 \fl p_{k|k_0}(\x,t|\x_0)=\Pi_{kk_0}(t)p(\x,t|\x_0),\quad J_{k_0}(\y|\x_0,t)=\Pi_{-1,k_0}(t)J(\y|x_0,t).
 \end{equation}
 Laplace transforming the renewal equation (\ref{2Dren}) using the convolution theorem then gives
 \begin{eqnarray}
\label{2DLT}
\fl \wrho^{(\epsilon)}_{k|k_0}(\x,s|\x_0)=\p_{k|k_0}(\x,s|\x_0)+\int_{\partial \Omega}   \wrho^{(\epsilon)}_{k|-1}(\x,s|\a_{\epsilon}(\y)) \J_{k_0}(\y,s|\x_0)d\y,
 \end{eqnarray}
 with
 \begin{eqnarray}
 \label{pk2D}
&\p_{k|k_0}(\x,s|\x_0)=\frac{1}{2}\bigg (\G(\x,s|\x_0)+kk_0\G(\x,s+2\alpha|\x_0)\bigg ),
\end{eqnarray}
and (for $\x_0\notin \partial \Omega$)
 \begin{eqnarray}
\fl  \widetilde{J}_{k_0}(\y,s|\x_0)&= \frac{1}{2}\bigg (\J(\y,s|\x_0)-k_0\J(\y,s+2\alpha|\x_0)\bigg )\nonumber \\
 \fl &=-\frac{D}{2}\bigg ( {\bm \nabla} \G(\y,s|\x_0)-k_0{\bm \nabla} \G(\y,s+2\alpha|\x_0)\bigg )\cdot \n(\y)  .
 \label{J2D}
  \end{eqnarray}
  Here $\G(\x,s|\x_0)$ is the Green's function of the modified Helmholtz equation in $\Omega $:
 \begin{eqnarray}
    \label{G2D}
\fl D{\bm \nabla}^2 \G(\x,s|\x_0)-s\G(\x,s|\x_0)&=-\delta(\x-\x_0),\ \x \in \Omega ,\quad
 \G(\y,s|\x_0)=0 \quad \y \in \partial \Omega.
 \end{eqnarray}
 Since the surface $\partial \Omega$ is spatially extended for $d>1$, the corresponding renewal equation (\ref{2DLT}) involves a spatial integration with respect to points $\y \in \partial \Omega$. However, this can be dealt with using spectral methods.

The first step is to set $k_0=-1$ and $\x_0=\a_{\epsilon}(\z)$ in equation (\ref{2DLT}) with $\z \in \partial \Omega$, which leads to a Fredholm integral equation of the second kind:
 \begin{eqnarray}
\label{2DLT2}
\fl \wrho^{(\epsilon)}_{k|-1}(\x,s|\a_{\epsilon}(\z))=\p_{k|-1}(\x,s|\a_{\epsilon}(\z))+ \int_{\partial \Omega}   \wrho^{(\epsilon)}_{k|-1}(\x,s|\a_{\epsilon}(\y)) \J_{-1}(\y,s|\a_{\epsilon}(\z))d\y .
 \end{eqnarray}
 It is convenient to rewrite equation (\ref{2DLT2}) in a more compact form. Temporarily suppressing the dependence on $k,\x,s$, we define
 \begin{equation}
 \label{hu}
 h^{(\epsilon)}(\z)=\p_{k|-1}(\x,s|\a_{\epsilon}(\z)),\quad u^{(\epsilon)}(\z)=\wrho^{(\epsilon)}_{k|-1}(\x,s|\a_{\epsilon}(\z)),
 \end{equation}
 and introduce the linear operator $ \L^{(\epsilon)}$ such that for any function $u\in L^2(\partial \calM)$
 \begin{equation}
 \L^{(\epsilon)}[u](\z)= \int_{\partial \Omega}u(\y) \J_{-1}(\y,s|\a_{\epsilon}(\z))d\y .
 \end{equation}
 We then have
 \begin{eqnarray}
\label{2DLT2b}
u^{(\epsilon)}(\z)=h^{(\epsilon)}(\z)+ \L^{(\epsilon)}[  u^{(\epsilon)}](\z) ,
 \end{eqnarray}
 which is a Fredholm integral equation of the second kind for $L^2$- integrable functions on $\partial \calM$. Since $\partial \calM$ is a compact domain, it follows that the linear operator  $\L^{(\epsilon)}$ has a discrete spectrum. Hence, one way to formally solve equation (\ref{2DLT2b}) is to perform an eigenfunction expansion of the functions $u^{(\epsilon)}(\z)$ and $h^{(\epsilon)}(\z)$. Here we will use such a spectral decomposition to establish that there is a well-defined solution in the limit $\epsilon \rightarrow 0$.
 
In order to take the limit $\epsilon \rightarrow 0$ we introduce the Taylor expansion
 \begin{equation}
 \label{Jeps}
 \fl \J_{-1}(\y,s|\a_{\epsilon}(\z))= \J_{-1} (\y,s| \z)- \epsilon \left . {\bm \nabla }_{\a} \J_{-1}(\y,s|\a)\right |_{\a=\z}\cdot \n(\z)+O(\epsilon^2) .
 \end{equation}
Note that care has to be taken in defining the boundary flux when the initial condition is also on the boundary, see Ref.  \cite{Bressloff23} and Appendix A. In particular, if $\z\in \partial \Omega$ then $\J(\x,s|\z)\equiv D{\bm \nabla} \G(\x,s|\z)\cdot \n(\x)=0$ for all $\x\not \in \partial \Omega$ due to the Dirichlet boundary condition $G(\x,s|\z)=0$. On the other hand, $ \J(\y,s| \z)=\overline{\delta}(\y-\z)$ for $\y,\z\in \partial \Omega$, where the Dirac delta function is restricted to the surface $\calU$.
 Combining equations (\ref{2DLT2b}) and (\ref{Jeps}) then yields
   \begin{eqnarray}
\label{2DLT3}
\fl u^{(\epsilon)}(\z)
&= h^{(\epsilon)}(\z) + \int_{\partial \Omega}  u^{(\epsilon)}(\y) \bigg [ \overline{\delta}(\z-\y)+ \frac{\epsilon D}{2}{\bm \nabla }_{\z} \bigg ({\bm \nabla }_{\y} \G(\y,s|\z)\cdot \n(\y)\bigg )\cdot \n(z) \\
\fl &\hspace{3cm} + \frac{\epsilon D}{2} {\bm \nabla }_{\z} \bigg ({\bm \nabla }_{\y} \G(\y,s+2\alpha|\z)\cdot \n(\y)\bigg )\cdot \n(z)+O(\epsilon^2)\bigg ]d\y. \nonumber 
 \end{eqnarray}
 Dropping $O(\epsilon^2)$ terms, equation (\ref{2DLT3}) can be written in the form
  \begin{equation}
 \label{fL}
\L_{s,\alpha}[u^{(\epsilon)}]=\epsilon^{-1}h^{(\epsilon)},
 \end{equation}
where $\L_{s,\alpha}$ is given by the D-to-N operator 
 \begin{eqnarray}
\label{DtoN}
\fl \L_{s,\alpha}[u](\z) &=-\frac{D}{2}\n(\z)\cdot {\bm \nabla }_{\z} \int_{\partial \Omega} \bigg ({\bm \nabla }_{\y} \bigg [\G(\y,s|\z)+\G(\y,s+2\alpha|\z)\bigg ]\cdot \n(\y)\bigg )u(\y) d\y\nonumber 
\\\fl 
\end{eqnarray}
acting on the space $L^2(\partial \Omega)$. (Note that $\L_{s,0}$ plays an important role in the spectral decomposition of solutions to the classical Robin boundary value problem for diffusion in $\Omega$ \cite{Grebenkov19}), thus establishing another mathematical connection between stochastically-gated and Robin boundary conditions.) 
The D-to-N operator $\L_{s,\alpha}$ has a discrete spectrum on $\partial \Omega$. That is, there exist countable sets of eigenvalues $\mu_n(s,\alpha)$ and eigenfunctions $v_n(\z,s,\alpha)$ satisfying
\begin{equation}
\label{eig}
\L_{s,\alpha} v_n(\z,s,\alpha)=\mu_n(s,\alpha)v_n(\z,s,\alpha),\quad \z\in \partial \Omega.
\end{equation}
It can be shown that the eigenvalues are positive definite (for $\alpha >0$) and that the eigenfunctions form a complete orthonormal basis in $L^2(\partial \Omega)$ with the orthogonality condition
\begin{equation}
\int_{\partial \Omega} v_n^*(\z,s,\alpha)v_m(\z,s,\alpha)d\z=\delta_{m,n}.
\end{equation}
We can now solve equation (\ref{fL}) by introducing an eigenfunction expansion of $u^{(\epsilon)}$,
\begin{equation}
\label{eig2}
u^{(\epsilon)}(\z)=\sum_{m=0}^{\infty}u^{(\epsilon)}_m(s,\alpha)  v_m(\z,s,\alpha).
\end{equation}
(The solution picks up a dependence on the switching rate $\alpha$ through the operator $\L_{s,\alpha}$.)
Substituting equation (\ref{eig2})
into (\ref{fL}) and taking the inner product with the adjoint eigenfunction $v_n^*(\x,s,\alpha)$ yields 
\begin{equation}
 \label{spec2}
u^{(\epsilon)}_m(s,\alpha)= \frac{1}{\epsilon} \frac{{\mathcal H}^{(\epsilon)}_n(s,\alpha)}{\mu_n(s,\alpha)},\quad {\mathcal H}^{(\epsilon)}_n(s,\alpha)= \int_{\partial \Omega}v_n^*(\z,s,\alpha) h^{(\epsilon)}(\z)d\z.
\end{equation}
Combining with equations (\ref{hu}), we have the result (assuming we can reverse the order of summation and integrations)
\begin{eqnarray}
\label{cool}
\fl  \wrho^{(\epsilon)}_{k|-1}(\x,s|\a_{\epsilon}(\y)) =\frac{1}{\epsilon}\int_{\partial \Omega}\bigg [\sum_{n=0}^{\infty}  \frac{v_n(\y,s,\alpha)v_n^*(\z,s,\alpha)}{\mu_n(s,\alpha)}\bigg ] \p_{k|-1}(\x,s|\a_{\epsilon}(\z))d\z .
 \end{eqnarray}

We can now substitute (\ref{cool}) into the renewal equation (\ref{2DLT}) and safely take the limit $\epsilon \rightarrow 0$ since $\p_{k|-1}(\x,s|\a_{\epsilon}(\z))=O(\epsilon)$. If we also sum equation (\ref{2DLT}) over $k$ and integrate with respect to $\x\in \Omega$, then we obtain the corresponding renewal equation for the Laplace transformed survival probability
 \begin{equation}
  \calQ_{k_0}(\x_0,s):=\int_{\Omega}\sum_{k=\pm 1}\wrho_{k|k_0}(\x,s|\x_0) d\x, 
  \end{equation}
  which takes the form
  \begin{eqnarray}
\label{2DLTQ}
\fl &\calQ_{k_0} (\x_0,s)\\
\fl &=\Q(\x_0,s)\ +\int_{\partial \Omega} d\y\int_{\partial \Omega}d\z \bigg [ \sum_{n=0}^{\infty}  \frac{v_n(\y,s,\alpha)v_n^*(\z,s,\alpha)  }{\mu_n(s,\alpha)} \bigg ]\bigg [\lim_{\epsilon \rightarrow 0}\frac{\Q(\a_{\epsilon}(\z),s)}{\epsilon}\bigg ]\J_{k_0}(\y,s|\x_0) .\nonumber
 \end{eqnarray}
Here $Q(\x_0,t)$ is the survival probability with respect to adsorption and
 \numparts
 \begin{eqnarray}
 \label{Qdef2D} 
  \Q(\x_0,s)  
 &:=\int_{\Omega} \p(\x,s|\x_0)d\x=\frac{1}{s}\bigg [1-\int_{\partial \Omega}\J(\y',s|\x_0)d\y'\bigg ].
 \end{eqnarray} 
 \endnumparts
 It follows that
 \begin{eqnarray}
 \fl \Q(\z-\epsilon \n(\z),s) & =\frac{1}{s}\bigg [1-\int_{\partial \Omega}\J(\y',s|\z)d\y'+\epsilon \int_{\partial \Omega}{\bm \nabla}_{\z}J(\y',s|\z)\cdot \n(\z) d\y'+O(\epsilon^2)\bigg ]\nonumber \\
\fl  &=-\frac{\epsilon D}{s}{\bm \nabla}_{\z}\bigg (\int_{\partial \Omega}{\bm \nabla}_{\y'} \G(\y',s|\z)\cdot \n(\y') d\y ' \bigg ) \cdot \n(\z)+O(\epsilon^2).
 \end{eqnarray}
 Substituting into equation (\ref{2DLTQ}), we have the following double-integral:
 \begin{eqnarray}
{\mathcal I}&=-\frac{D}{s}\int_{\partial \Omega}d\z \int_{\partial \Omega}d\y' v_n^*(\z,s,\alpha) {\bm \nabla}_{\z}\bigg ({\bm \nabla}_{\y'} \G(\y',s|\z)\cdot \n(\y')\bigg )\cdot \n(\z) \nonumber\\
&=\frac{1}{s} \int_{\partial \Omega} d\y'\, {\mathbb L}^*_{s,0} v_n^*(\y',s,\alpha)  .
\end{eqnarray}
Hence, equation (\ref{2DLTQ}) can be rewritten as
  \begin{eqnarray}
\label{2DLTQ2}
\fl \calQ_{k_0} (\x_0,s)&=\Q(\x_0,s) +\frac{1}{s} \sum_{n=0}^{\infty}\frac{1}{\mu_n(s,\alpha)}\bigg [ \int_{\partial \Omega} d\y\, v_n(\y,s,\alpha)\J_{k_0}(\y,s|\x_0)\bigg ]\\
\fl &\quad  \hspace{5cm}\times  \bigg [  \int_{\partial \Omega} d\y'\, {\mathbb L}^*_{s,0} v_n^*(\y',s,\alpha)  \bigg ].\nonumber
 \end{eqnarray}

\subsection{Finite interval revisited} Let us briefly return to the example of a finite interval $\Omega =(0,L)$ for which $\partial \calU=\{0,L\}$. The corresponding Dirichlet Green's function is given by equation (\ref{G1D}) and the D-to-N operator is a $2\times 2$ matrix ${\bf L}(s,\alpha)$ with elements
\begin{eqnarray}
\fl L_{ab}(s,\alpha) =  -\frac{D}{2}(-1)^a(-1)^b\left .\partial_{x}\partial_{x'} [\G(x',s|x)+\G(x',s+2\alpha|x)\right |_{x'=aL,x=bL},
\end{eqnarray}
for $a,b=0,1$. (The factors of $-1$ ensure that the fluxes are directed towards the exterior of the interval.) We thus find that $L_{ab}(s,\alpha)=-(-1)^a[{\bf m}(s)+{\bf m}(s+2\alpha)]_{ab}$, see equation (\ref{M}). Moreover, in
equation (\ref{2DLTQ}) we have the simplifications 
\[\int_{\partial \calU}d\y \, \int_{\partial \calU}d\z \rightarrow \sum_{a=0,1}\sum_{b=0,1},\]
and
\[ \fl \sum_{n=0}^{\infty}  \frac{v_n(\y,s,\alpha)v_n^*(\z,s,\alpha)  }{\mu_n(s,\alpha)}\rightarrow \sum_{n=0,1}\frac{v_n(aL,s,\alpha)v_n^*(bL,s,\alpha)  }{\mu_n(s,\alpha)}\equiv [{\bf L}(s,\alpha) ]_{ab}^{-1}.\]
We thus recover equation (\ref{ren2gateQ}).

\subsection{Stochastically gated sphere}

\begin{figure}[t!]
\centering
\includegraphics[width=8cm]{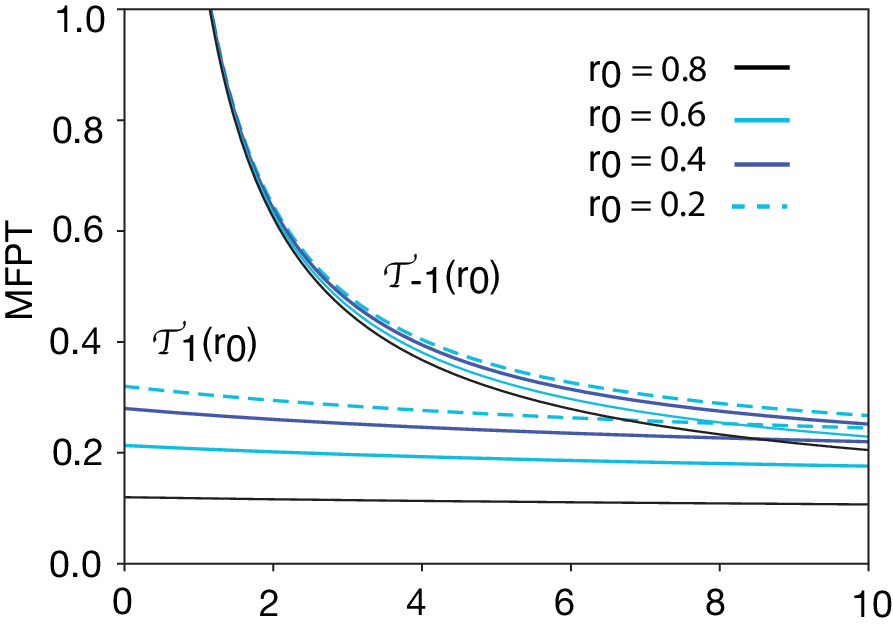} 
\caption{Brownian particle in a stochastically gated sphere. Plots of the MFPTs $\calT_{\pm 1}(r_0)$ as a function of the switching rate $\alpha$ for various values of the initial radial position $r_0$. Other parameter values are $D=1$ and $R=1$.}
\label{fig6}
\end{figure}

One example where the spectral decomposition of the D-to-N operator $\L(s,\alpha)$ can be calculated explicitly is on the surface $\partial \Omega$ of a sphere. Let $\Omega=\{\x\in \R^3,\, 0 <  |\x| <R\}$ so that $\partial \Omega= \{\x\in \R^3,\,  |\x| =R\}$. The rotational symmetry of $\Omega$ means that if $\L_{s,\alpha}$ is expressed in spherical polar coordinates $(r,\theta,\phi)$, then the eigenfunctions are given by spherical harmonics, and are independent of the Laplace variable $s$ (and hence $\alpha$):
\begin{equation}
v_{nm}(\theta,\phi)=\overline{v}_{nm}(\theta,\phi)=\frac{1}{R} Y_n^m(\theta,\phi),\quad n\geq 0, \ |m|\leq n,
\end{equation}
where
\begin{equation}
Y_n^m(\theta,\phi)=\sqrt{\frac{2n+1}{4\pi} \frac{(n-m)!}{(n+m)!}}P_n^m(\cos\theta)\e^{im\phi},
\end{equation}
and $P_l^m(\theta)$ is an associated Legendre polynomial.
From orthogonality, it follows that the adjoint eigenfunctions are
\begin{equation}
v^*_{nm}(\theta,\phi)=\overline{v}_{nm}^*(\theta,\phi)=(-1)^m\frac{1}{R} Y_n^{-m}(\theta,\phi).
\end{equation}
(Note that eigenfunctions are labeled by the pair of indices $n,m$.) We can now use the principle of superposition to write down the eigenvalues $\mu_n(s,\alpha)$, given the corresponding eigenvalues for $\alpha=0$ \cite{Grebenkov19}:
\begin{equation}
\fl \mu_n(s,\alpha)=\frac{1}{2}\bigg [\sqrt{\frac{s}{D}}\frac{i_n'(\sqrt{s/D} R)}{i_n(\sqrt{s/D} R)}+\sqrt{\frac{s+2\alpha}{D}}\frac{i_n'(\sqrt{[s+2\alpha]/D} R)}{i_n(\sqrt{[s+2\alpha]/D} R)}\bigg ],
\end{equation}
where
\begin{equation}
i_n(z)=\sqrt{\frac{\pi}{2z}} I_{n+1/2}(z).
\end{equation}
is the modified spherical Bessel function of the first kind. 
Since the $n$th eigenvalue is independent of $m$, it has a multiplicity $2n+1$. Hence, rewriting equation (\ref{2DLTQ2}) in spherical polar coordinates with $\x_0=(r_0,\theta_0.\phi_0)$ and $r<R$, we obtain the decomposition
\begin{eqnarray}
\label{2DLTQsp}
\fl \calQ_{k_0} (\x_0,s)&=\Q(\x_0,s) +\frac{R^2}{s}  \int {\mathcal D}(\theta,\phi)\int {\mathcal D}(\theta',\phi')\sum_{n=0}^{\infty} \sum_{m=0}^n \frac{\mu_n(s,0)}{\mu_n(s,\alpha)}\\
\fl &\qquad\hspace{3cm}  \times  (-1)^m Y_n^m(\theta,\phi) Y_n^{-m}(\theta',\phi') \J_{k_0}(R,\theta,\phi,s|\x_0),\nonumber 
\end{eqnarray}
where
\begin{equation}
\int {\mathcal D}(\theta,\phi):=\int_0^{\pi}\sin \theta d\theta \int_{-\pi}^{\pi}d\phi .
\end{equation}
Using the identity
\begin{equation}
\int{\mathcal D}(\theta',\phi')Y_n^{-m}(\theta',\phi') =\delta_{n,0}\delta_{m,0}\sqrt{4\pi},
\end{equation}
we obtain the simplified result
\begin{eqnarray}
\label{2DLTQsp2}
\calQ_{k_0} (\x_0,s)&=\Q(\x_0,s) +\frac{R^2}{s}  \frac{\mu_0(s,0)}{\mu_0(s,\alpha)} \int {\mathcal D}(\theta,\phi) \J_{k_0}(R,\theta,\phi,s|\x_0). 
\end{eqnarray}
Finally, noting from equation (\ref{J2D}) that
\begin{eqnarray}
\fl \J_{k_0}(R,\theta,\phi,s|\x_0)&=\J(R,\theta,\phi,s|\x_0)-k_0 \J(R,\theta,\phi,s+2\alpha|\x_0)\nonumber\\
\fl &=-\left . D\partial_r \bigg (\G(r,\theta,\phi,s|\x_0)-k_0 \G(r,\theta,\phi,s+2\alpha|\x_0)\bigg )\right |_{r=R},
\end{eqnarray}
and using the second identity \cite{Grebenkov19}
 \begin{eqnarray}
\fl -DR^2\int {\mathcal D}(\theta,\phi) Y_n^m(\theta,\phi) \partial_r\G(R,\theta,\phi,s|\x_0)
=Y_n^m(\theta_0,\phi_0) \frac{i_n(\sqrt{s/D} r_0)}{i_n(\sqrt{s/D} R)}
\end{eqnarray}  
for $\x_0=(r_0,\theta_0,\phi_0)$ and $r_0<r$, we see that
\begin{equation}
\fl  \int {\mathcal D}(\theta,\phi) \J (R,\theta,\phi,s|\x_0)=\frac{1}{R^2}\frac{i_0(\sqrt{s/D} r_0)}{i_0(\sqrt{s/D} R)},\quad i_0(z)=\frac{\sinh z}{z}.
 \end{equation}
 Hence,
\begin{eqnarray}
\label{2DLTQsp3}
\fl \calQ_{k_0} (\x_0,s)&=\Q(\x_0,s) +\frac{1}{2s}  \frac{\mu_0(s,0)}{\mu_0(s,\alpha)}\bigg [ \frac{i_0(\sqrt{s/D} r_0)}{i_0(\sqrt{s/D} R)}-k_0\frac{i_0(\sqrt{[s+2\alpha]D} r_0)}{i_0(\sqrt{[s+2\alpha]/D} R)}\bigg ],
\end{eqnarray}
with
\begin{eqnarray}
 \fl  \Q(\x_0,s)  
 &=\frac{1}{s}\bigg [1-R^2\int {\mathcal D}(\theta',\phi')\J(R,\theta',\phi',s|\x_0)\bigg ]=\frac{1}{s}\bigg [1- \frac{i_0(\sqrt{s/D} r_0)}{i_0(\sqrt{s/D} R)}\bigg ].
 \end{eqnarray} 
We now take the limit $s\rightarrow 0$ in equation (\ref{2DLTQsp3}) using the small-$z$ expansion $i_0(z)=1+z^2/6+O(z^4)$. This yields
\begin{eqnarray}
\fl \calT_{k_0}(\x_0)=\frac{R^2-r_0^2}{6D}+\frac{R}{3D}\bigg [ \sqrt{\frac{D}{2\alpha}}\frac{i_0(\sqrt{2\alpha/D} R)}{i_0'(\sqrt{2\alpha /D} R)}\bigg ]\bigg [1-k_0 \frac{i_0(\sqrt{2\alpha/D} r_0)}{i_0(\sqrt{2\alpha/D} R)}\bigg ].
\end{eqnarray}
Similar to previous examples, we find that in the slow switching limit
\begin{equation}
\lim_{\alpha \rightarrow 0} \calT_{-1}(x_0)=\infty,\quad \lim_{\alpha \rightarrow 0} \calT_{1}(x_0)=2\frac{R^2-r_0^2}{6D}.\end{equation}
In particular, the MFPT for $k_0=1$ is double the classical result for a totally absorbing sphere (no gate). In Fig. \ref{fig6} we show example plots of the MFPTs $\calT_{\pm 1}(r_0)$ as a function of the switching rate $\alpha$ for various values of the initial radial position $r_0$. We see that the MFPTs are monotonically decreasing functions of $\alpha$ and $r_0$.

 \section{Discussion}
 
 In this paper we developed a renewal theory for diffusion in a domain with one or more stochastically gated boundaries. We assumed that a boundary randomly switches between a closed (reflecting) and an open (absorbing) state according to a two-state Markov chain. The gating could be due to physical changes in the diffusing particle or the boundary itself. Through a number of examples, we showed how to construct a renewal equation that relates the joint probability density for particle position and the state of a gate (or multiple gates) to the marginal probability density and FPT density in the case of totally absorbing (non-gated) boundaries; the latter are typically easier to calculate then solving the forward Kolmogorov equation for the fully gated system. The renewal equation effectively decomposes sample paths into an alternating sequence of bulk diffusion and instantaneous adsorption/desorption events, which is terminated when adsorption coincides with an open gate. 
 This formulation has a number of other advantages over traditional methods. First, the renewal equation has an intuitive causal structure in the time domain that makes it relatively straightforward to construct for different geometric configurations. Second, explicitly separating absorption and adsorption/desorption events provides a powerful mathematical framework for incorporating more general features such as partial adsorption and boundary-induced resetting. Third, the spectral theory of D-to-N operators can be used to solve the renewal equation in higher spatial dimensions, which involves a surface integral over the spatially extended target boundary.
 
There is one limitation of the renewal theory for stochastically gated targets that does not occur for adsorption/desorption processes \cite{Bressloff25a,Bressloff25b}. In the latter case, the Laplace transformed renewal equations can still be solved explicitly when the waiting time density $\phi(\tau)$ for the time $\tau$ attached to a sticky wall is a non-exponential function. That is, one can construct non-Markovian models of desorption/absorption. On the other hand, the renewal equation (\ref{renewal}), for example, involves the product $p_{k|k_0}(x,t|x_0):=\Pi_{kk_0}(t)p(x,t|x_0)$, where $\Pi_{kk_0}(t)$ is the transition matrix for the switching gate. We relied on the fact that in the case of Markovian switching, $\Pi_{kk_0}(t)$ consists of a linear combination of exponential functions of time, so that we can express the Laplace transform $\p_{k|k_0}(x,s|x_0)$ in terms of the corresponding marginal density $\p(x,s|x_0)$. Incorporating non-Markovian models of a switching gate is more of a challenge.

\begin{figure}[t!]
\centering
\includegraphics[width=12cm]{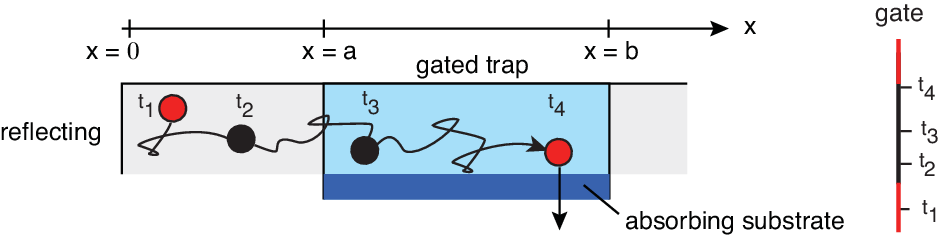} 
\caption{1D diffusion in the half-line $\Omega=[0,\infty)$ with a stochastically gated trap $\calU=[a,b]$. The boundary $x=0$ is totally reflecting. The particle randomly switches between a reactive and non-reactive state, and can only be absorbed within $ \calU$ when in the reactive state, otherwise it continues to diffuse. Alternatively, the absorbing substrate could be physically gated. (The 2D representation is for visualisation purposes.)}
\label{fig7}
\end{figure}

Another major challenge is extending the renewal theory to the case of stochastically-gated interior traps rather than target boundaries. Consider, for example, a stochastically gated trap $\calU=[a,b]$ in the interior of the half-line $\Omega =[0,\infty)$, see Fig. \ref{fig7}. The particle can now freely enter and exit $ \calU$ and is absorbed within $\calU$ as soon as the gate opens. We would like to construct a renewal equation that relates the joint propagator $\rho_{k|k_0}$ to a marginal propagator $p$ for which the substrate is totally absorbing (non-gated). However, the latter would only represent sample paths that are restricted to the exterior of $\calU$ since as soon as a particle reaches $\partial \calU$ it would be removed.  Hence, this would exclude sample paths that start within the interior of the trap when the gate is closed. One way to proceed would be to treat $\calU$ as a partially adsorbing substrate with reactivity $\kappa_0$ along analogous lines to 
Fig. \ref{fig4}(a) so that the particle only detects the state of the gate after it is adsorbed. Alternatively, one could treat the open substrate as partially absorbing along analogous lines to Fig. \ref{fig4}(b) and use the Green's function method of Szabo {\em et al.} \cite{Szabo84,Spouge96}. However, both approaches lead to integral equations involving spatial integrals over the trapping region so that the analysis is considerably more involved, particularly with regards retaking the limit $\kappa_0\rightarrow \infty$.
Recently Sher {\em et al.} \cite{Scher24} have developed a method that avoids the occurrence of spatial integrals. They derive a recursive relation between the FPT density for absorption to the corresponding FPT density for adsorption in the absence of gating by conditioning on whether the particle is located within or outside the trap when the gate reopens. This yields exact expressions for the Laplace transform of the FPT density for a 1D trap. However, extending the latter method to higher-dimensional traps is itself non-trivial.

   \begin{figure}[b!]
  \centering
  \includegraphics[width=12cm]{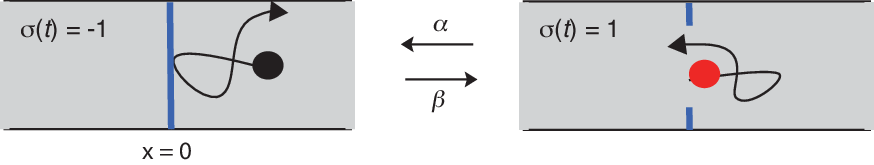}
  \caption{Brownian particle in $\R $ with a stochastically gated interface at $x=0$. The gate is open (permeable) if $\sigma(t)=1$ and closed (impermeable) if`$\sigma(t)=-1$. Transitions $\sigma(t)\rightarrow -\sigma(t)$ occurs according to a two-state Markov chain with transition rates $\alpha,\beta$. (The 2D representation is for illustrative purposes only.)}
  \label{fig8}
\end{figure}

Finally, note that another class of problems where the renewal formulation developed in this paper can be applied is diffusion across a stochastically gated interface within the interior of a domain. Consider the example of BM in $\R$ with an interface at $x=0$, see Fig. \ref{fig8}. Following along analogous lines to section 2, one can construct a renewal equation that sews together a sequence of BMs on the half-line with a totally adsorbing boundary at $x=0$. Now, however, each time the particle is adsorbed, the stochastic process is immediately restarted according to the following modified rule: if the gate is closed then diffusion restarts on the same side of the interface, whereas if the gate is open then diffusion restarts on either side of the interface with equal probability. Again, in order to ensure that diffusion restarts in a state that avoids immediate re-absorption, the particle is instantaneously shifted a distance $\epsilon$ from the origin whenever it reaches the interface. The renewal theory for stochastically gated interfaces is developed in a companion paper \cite{Bressloff26}. For a corresponding theory of semi-permeable interfaces based on so-called snapping out BM see Ref. \cite{Bressloff23}.
 
 .

 
 \setcounter{equation}{0}
\renewcommand{\theequation}{A.\arabic{equation}}
\section*{Appendix A: Higher-dimensional boundary flux}

 \begin{figure}[b!]
\centering
  \includegraphics[width=6cm]{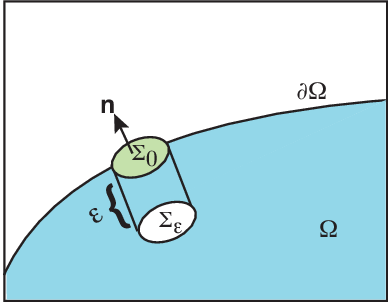}
  \caption{Cylinder construction at a target surface $\partial \calU$. See text for details.}
  \label{fig9}
\end{figure}

 In section 4, we displaced a particle hitting a point $\y \in \partial \Omega$ on a closed surface according to $\y\rightarrow \y-\epsilon \n(\y)$. Here we provide further details regarding the flux on the boundary. In order to proceed we consider the forward Kolmogorov equation in Laplace space:
\numparts 
\begin{eqnarray}
\label{pdifflocLTa}
	& D\nabla_{\x}^2 \p(\x,s|\x_0) -s\p(\x,s|\x_0)=-\delta(\x-\x_0),\, \x,\x_0 \in \Omega .\\
& \p(\x,s|\z) =0 \mbox{ for } \z\in \partial \Omega.
\label{pdifflocLTb}
	\end{eqnarray}
	\endnumparts 
Consider a small cylinder ${\mathcal C}(\epsilon,\sigma)$ of uniform cross-section $\sigma$ and length $\epsilon$ as shown in Fig. \ref{fig9}. For sufficiently small $\sigma$, we can treat $\Sigma_{0}\equiv {\mathcal C}(\epsilon,\sigma)\cap \partial \Omega$ as a planar interface with outward normal $\n(\y)$ such that the axis of ${\mathcal C}(\epsilon,\sigma)$ is aligned along $\n(\y)$. Given the above construction, we integrate equation (\ref{pdifflocLTa}) with respect to all $\x\in  {\mathcal C}(\epsilon,\sigma)$ and use the divergence theorem:
 \begin{eqnarray}
&\int_{\Sigma_{0}} \nabla \p(\y',s|\x_0) \cdot \n(\y') d\y' - \int_{\Sigma_{\epsilon}} \nabla \p(\y',s|\x_0)\cdot \n(\y') d\y' \nonumber \\
&\quad \sim \frac{1}{D}\int_{\calC} [s\p(\x,s|\x_0)-\delta(\x-\x_0)] d\x,
\label{cyl}
 \end{eqnarray}
 where $\Sigma_{\epsilon}$ denotes the flat end of the cylinder within $\Omega $. If $\x_0$ is in the bulk domain, then taking the limits $\epsilon,\sigma \rightarrow 0$ shows that the flux is continuous as it approaches the boundary, since the right-hand side of equation (\ref{cyl}) vanishes. On the other hand, if $\x_0=\z \in\partial \Omega$ then taking the limits $\epsilon,\sigma \rightarrow 0$ gives
 \begin{eqnarray}
\fl  &D  \nabla \p(\y,s|\z) \cdot \n(\y)-\lim_{\epsilon \rightarrow 0^+} D \nabla \p(\y-\epsilon \n(\y),s|\z)\cdot \n(y) =-\overline{\delta}(\y-\z),
 \end{eqnarray}
 where $\overline{\delta}$ is the Dirac delta function for points on $\partial \Omega$ such that for any continuous function $f: \calU\rightarrow \R$ we have
 $\int_{\partial \Omega} f(\y)\overline{\delta}(\y-\z)d\y = f(\z)$. Finally, noting that the second flux term on the left-hand side vanishes, since  $\p(\y+\epsilon \n(\y),s|\z)$ for $\epsilon >0$, we deduce that
 \begin{equation}
 \label{crucial}
 J(\y,s|\z)\equiv -D  \nabla \p(\y,s|\z) \cdot \n(\y)=\overline{\delta}(\y-\z), \quad \y,\z\in \partial \Omega.
 \end{equation}

 \enlargethispage{20pt}








\end{document}